\documentclass[12pt,ta4paper]{article}

\usepackage[colorlinks,pdftex,citecolor=blue,urlcolor=blue]{hyperref}
\usepackage{lineno}
\usepackage{amsmath}
\usepackage{textcomp} 

\usepackage{natbib}
\usepackage{graphicx}
\usepackage{amssymb}
\usepackage{bm}

\usepackage{a4wide}
\usepackage[T1]{fontenc}
\usepackage[utf8]{inputenc}
\usepackage{tgschola}

\title{
What limits the number of observations that can be effectively assimilated by EnKF?
\footnote{submitted to QJRMS}
}
\author{
  Daisuke Hotta
   \thanks{Meteorological Research Institute, Japan Meteorological Agency, Tsukuba, Japan}
   \thanks{Numerical Prediction Division, Japan Meteorological Agency, Tokyo, Japan}
   \thanks{European Centre for Medium-range Weather Forecasts, Reading, UK}
  \and
  Yoichiro Ota\footnotemark[3]
}

\begin{document}
\maketitle

\begin{abstract}
The ability of ensemble Kalman filter (EnKF) algorithms to extract
information from observations is analyzed with the aid of the concept of
the degrees of freedom for signal (DFS). A simple mathematical argument
shows that DFS for EnKF is bounded from above by the ensemble size,
which entails that assimilating much more observations than the ensemble
size automatically leads to DFS underestimation. Since DFS is a trace of
the posterior error covariance mapped onto the normalized observation
space, underestimated DFS implies overconfidence (underdispersion) in
the analysis spread, which, in a cycled context, requires covariance
inflation to be applied.
The theory is then extended to cases where covariance localization
schemes (either B-localization or R-localization) are applied to show
how they alleviate the DFS underestimation issue.
These findings from mathematical argument are demonstrated with a
simple one-dimensional covariance model.
Finally, the DFS concept is used to form speculative arguments about how
to interpret several puzzling features of LETKF previously reported in
the literature such as why using less observations can lead to better
performance, when optimal localization scales tend to occur, and why
covariance inflation methods based on relaxation to prior information
approach are particularly successful when observations are
inhomogeneously distributed.
A presumably first application of DFS diagnostics to a quasi-operational
global EnKF system is presented in Appendix.
%
\end{abstract}


\newcommand{\ie}{\textit{i.e.}}
\newcommand{\tr}{\mathrm{tr}\ }


\section{Introduction}

The number of observations that are available for operational numerical
weather prediction (NWP) systems has undergone dramatic increase over
the last several decades. This increasing trend, largely driven by
advances in remote-sensing technology, is envisaged to continue in the
near future thanks to the incoming meteorological Big Data such as
measurements of phased-array weather radars \citep{Miyoshi2016} and
satellite-based hyper-spectral soundings. A challenge in data
assimilation (DA) development that is becoming increasingly relevant
today is thus to effectively exploit the ever-increasing amount of
observations that are becoming denser and more frequent. More
specifically, we need to be able to {\it extract as much information as
possible} from such observations. This is not an easy task, even for
advanced DA algorithms such as ensemble Kalman filter (EnKF),
partly because these algorithms are built upon the assumption that the
number of observations to be assimilated is orders of magnitude
smaller than the degrees of freedom of the state space --- an assumption
that was perfectly legitimate when these algorithms were devised several
decades ago but may need to be revisited given the current explosive
increase in the volume of observations.

In this context, it is important to quantify the amount of information
that a DA system can extract from observations. Several criteria for
quantifying the amount of information have been proposed in the past,
and degrees of freedom for signal (DFS), or information content, is an
example of such measures.
The theory of DFS was developed in the statistics literature for general
inverse problems \citep{Wahba1995,Rodgers2000} and has been adapted to
address many problems that arise in NWP \citep[e.g.][]{PurserHuang1993,Rabier2002,Bocquet2011}.
DFS is defined for linear Gaussian least-square DA schemes as the trace
of the ``influence matrix'' $(\mathbf{S}=(\mathbf{HK})^T)$ where
$\mathbf{H}$ and $\mathbf{K}$ denote, respectively, the Jacobian of the
observation operator and the Kalman gain matrix (see Section 2 for
details). For DA algorithms suitable for systems with large state-vector
dimensions like 4Dvar or EnKF, the Kalman gain $\mathbf{K}$ is usually
not explicitly constructed, so evaluation of DFS is not
straightforward. Several ways to approximately compute DFS for
variational DA systems at a feasible computational cost have been
proposed \citep{Cardinali2004,Fisher2003,Chapnik2006,Lupu2011} and DFS
has now become a standard diagnostics for assessing the impact of
different instruments or observing platforms for 4DVar-based operational
DA systems.

In the case of EnKF, \cite{Liu2009} showed that DFS can be easily
computed as long as the analysis perturbations projected onto
observation space are available. However, in contrast to variational DA
systems, to the authors' best knowledge, DFS has not yet been applied to
operational EnKF-based DA systems.

As we report in Appendix, we have applied DFS diagnostics to the
quasi-operational version of Japan Meteorological Agency (JMA)'s global
EnKF-based DA system to examine how much information our analysis
extracts from each type of observations. The diagnostics revealed,
intriguingly, that, whilst our EnKF system is extracting reasonable
amount of information from relatively sparse observations such as
conventional ground-based observations (SYNOP) and radiosondes (TEMP),
showing per-obs DFS comparable to that from 4DVar, it retrieves far less
information from dense observations such as the satellite radiances
(hyper-spectral soundings from IASI and AIRS in particular), with DFS an
order of magnitude smaller than in 4DVar. The theoretical arguments and
their demonstrations using a simple covariance model presented in this
manuscript grew out from our attempt to understand the reason behind
this problem. As we will show later, an argument based on DFS allows us
to clearly illustrate the relevance of the ensemble size (or the
effective rank of the background error covariance matrix) in effectively
extracting observational information in scenarios where a large volume
of observations are available.

The purpose of this paper is to analyze the properties of Ensemble
Transform Kalman filter \citep[ETKF][]{Bishop2001,Wang2004} and its
local variant \citep[LETKF][]{Hunt2007} in light of the concept of DFS.
We will show, without using anything beyond elementary linear algebra,
that DFS can be used to quantitatively describe the well-known (but
vaguely defined) ``rank deficiency'' issue that an EnKF system suffers
when the ensemble size is not sufficient. The theory developed here not
only has direct relevance to the important question of how many ensemble
members we need to effectively assimilate a given set of observations,
but also bares some interesting implication on covariance localization
and inflation --- the two crucial components of EnKF without which the
algorithm do not work but are often subject to manual tuning due to our
lack of understanding.

The rest of the paper is structured as follows: Section \ref{sec:dfs}
reviews the theory of DFS and shows how it is related to the singular
values of the ``observability matrix''
$\mathbf{R}^{-1/2}\mathbf{HB}^{1/2}$ (\ie, the square root of the
background error covariance matrix measured in observation space
normalized by the inverse square root of the observation errror
covariance matrix). Section \ref{sec:dfs-ens} then applies this theory
to ETKF and proves that the DFS that ETKF can attain is limited from
above by the ensemble size, which entails that the analysis becomes
suboptimal whenever the ensemble size is smaller than the DFS that
should be attained by an optimal analysis. Section \ref{sec:dfs-ens}
also discusses the implication of this DFS underestimation on covariance
localization and inflation. Section \ref{sec:exp} illustrates the
findings given in the preceding section through a series of idealized
experiments using a simple one-dimensional covariance model. Section
\ref{sec:discussion} provides several speculative discussions as to how
the logic developed in this paper can be used to interpret some puzzling
(or counter-intuitive) results previously reported in the literature,
followed by conclusions in Section \ref{sec:conclusions}.

\section{Degrees of Freedom for Signal (DFS)}
\label{sec:dfs}
\subsection{Brief review of DFS}

DFS is a measure of how much information an analysis has retrieved from
observations \citep[e.g.,][]{Rodgers2000}. For a linear Gaussian DA
scheme, DFS is defined as the trace of the influence matrix
$\mathbf{K}^T\mathbf{H}^T$:
\begin{align}
 \mbox{DFS} = \tr \mathbf{K}^T\mathbf{H}^T = \tr \mathbf{HK} \label{eq:dfs-def}
\end{align}
where $\tr$ denotes the trace of a matrix, $\mathbf{H}$ denotes the
linear observation operator and $\mathbf{K}$ denotes the Kalman gain
matrix. Since the analysis projected onto the observation space is
\begin{align}
 \mathbf{y}^a = \mathbf{H}\mathbf{x}^a
 = \mathbf{H}\mathbf{x}^b
 + \mathbf{HK}\left(\mathbf{y}^o-\mathbf{H}\mathbf{x}^b\right)
\end{align}
where $\mathbf{x}^b$ denotes the background mean, DFS can also be
expressed as
\begin{align}
 \mbox{DFS}=\tr \left( \frac{\partial\mathbf{y}^a}{\partial \mathbf{y}^o} \right)
 = \sum_{i=1}^{p}\frac{\partial y^a_i}{\partial y^o_i} \label{eq:self-sensitivity}
\end{align}
where $p$ denotes the number of the assimilated observations. Equation
\ref{eq:self-sensitivity} allows us to interpret DFS as the sensitivity
of analysis with respect to observations. Notably, the total DFS can be
partitioned into contributions from each observation (indexed by $i$),
and the contribution from each observation takes the form of
``self-sensitivity.''

The nature of DFS is better understood if we work in the singular space
of the observability matrix $\mathbf{R}^{-1/2}\mathbf{HB}^{1/2}$
\citep{Johnson2005} where $\mathbf{R}$ and $\mathbf{B}$ denote,
respectively, the observation and background error covariance matrices.
In the canonical Kalman filter (KF), the analysis error covariance matrix
$\mathbf{A}$ and the Kalman gain $\mathbf{K}$ are related by
\begin{align}
 \mathbf{K} &= \mathbf{AH}^T\mathbf{R}^{-1} \label{eq:K-AHtRinv}
\end{align}
so that 
\begin{align}
\mathbf{H'AH'}^T
 := \mathbf{R}^{-1/2}\mathbf{HAH}^T\mathbf{R}^{-1/2} 
  = \mathbf{R}^{-1/2}\mathbf{HK}\mathbf{R}^{1/2} \label{eq:HAH-def}
\end{align}
where we have introduced the normalized observation operator
$\mathbf{H}':=\mathbf{R}^{-1/2}\mathbf{H}$ to avoid cluttered notation.
Recalling that the Kalman gain $\mathbf{K}$ can be also expressed as
\begin{align}
 \mathbf{K} = \mathbf{BH}^T\left(\mathbf{HBH}^T+\mathbf{R}\right)^{-1}, \label{eq:Kgain-def}
\end{align}
Equation \ref{eq:HAH-def} can be expressed as
\begin{align}
\mathbf{H'AH'}^T
=  \left(\mathbf{H'BH'}^T\right)
  \left\{\left(\mathbf{H'BH'}^T\right) + \mathbf{I}_p \right\}^{-1} \label{eq:HAH}
\end{align}
where $\mathbf{I}_p$ denotes the identity matrix of size $p\times p$.
Now, introducing singular decomposition on the observability matrix $\mathbf{H'B}^{1/2}$:
\begin{align}
 \mathbf{H'B}^{1/2}=\mathbf{U}\bm{\Sigma}^b\mathbf{V}^T \label{eq:svd}
\end{align}
the background error covariance matrix projected onto  the normalized
observation space $\mathbf{H'BH'}^T$ is eigendecomposed as 
\begin{align}
 \mathbf{H'BH'}^T=\mathbf{U}\bm{\Lambda}^b\mathbf{U}^T, \label{eq:eigen-HBH}
\end{align}
which, plugged-in to Equation \ref{eq:HAH}, yields the eigen-decomposed
expression of $\mathbf{H'AH'}^T$ as
\begin{align}
 \mathbf{H'AH'}^T=\mathbf{U}\bm{\Lambda}^a\mathbf{U}^T \label{eq:eigen-HAH}
\end{align}
where the diagonal matrix $\bm{\Lambda}^a$ is defined later (see
Equation \ref{eq:eigval-A}). Denoting the $i$-th diagonal element of the
matrix $\bm{\Sigma}^b$ by $\sigma^b_i$, which are the singular
values of the observability matrix, and assuming that they are sorted
from the largest to the smallest, the $i$-th diagonal element of the
diagonal matrix $\bm{\Lambda}^b=\bm{\Sigma}^b{\bm{\Sigma}^b}^T$ is
\begin{align}
\lambda^b_i=
 \begin{cases}
  {\sigma^b_i}^2   &\qquad (i \leq r) \\
  0                &\qquad  (i > r)
 \end{cases}
\end{align}
where $r$ is the rank of the observability matrix
$\mathbf{H'B}^{1/2}$. The eigenvalues $\lambda^a_i$ of
$\mathbf{H'AH'}^T$ are then related to the eigenvalues $\lambda^b_i$ of
$\mathbf{H'BH'}^T$ by
\begin{align}
 \lambda^a_i = \frac{\lambda^b_i}{\lambda^b_i+1}. \label{eq:eigval-A}
\end{align}

Equations \ref{eq:eigen-HBH}--\ref{eq:eigval-A} are very helpful in
understanding why DFS is called as such. From Equation
\ref{eq:eigen-HBH} we see that, in the space spanned by each column of
$\mathbf{U}$, the state can vary statistically independently in each
direction, and the background error has the variance of $\lambda^b_i$ in
each direction. Each direction corresponds to one ``degree of freedom''
since it can vary independently of each other. Equation
\ref{eq:eigen-HAH} means that, in this space, the error variance
$\lambda^b_i$ in the $i$-th direction is reduced by a factor of
$\lambda^a_i/\lambda^b_i$ by assimilating observations. From Equation
\ref{eq:eigval-A}, this factor can be expressed as
$\lambda^a_i/\lambda^b_i = 1/(\lambda^b_i+1)$. Now, if $\lambda^b_i$ is much
greater than one ($\lambda^b_i \gg 1$), the error variance in the $i$-th
direction is reduced by a large fraction ($\lambda^a_i/\lambda^b_i \ll
1$), meaning that the uncertainty in this direction is well constrained
by the observations. Such a direction can be considered as representing
one ``degree of freedom for signal.''  Conversely, if $\lambda^b_i$ is
close to zero ($\lambda^b_i \ll 1$), the error variance in that
direction is hardly reduced ($\lambda^a_i/\lambda^b_i \approx 1$),
meaning that this direction is virtually not observed at all. Such a
direction can be considered as representing one ``degree of freedom for
noise'' (or equivalently, zero ``degree of freedom for signal''). The
eigenvalue of $\mathrm{H'AH'}^T$, $\lambda^a_i$, has an interesting
property of approaching one if $\lambda^b_i$ is large
($\lambda^a_i\rightarrow 1$ as $\lambda^b_i\rightarrow\infty$; \ie, the
$i$-th direction represents a degree of freedom for signal) and
approaching zero if $\lambda^b_i$ is close to zero
($\lambda^a_i\rightarrow 0$ as $\lambda^b_i\rightarrow 0$; \ie, the
$i$-th direction represents a degree of freedom for noise). It is then
sensible to define the total ``degrees of freedom for signal'' (DFS) as
the sum of all $\lambda^a_i$'s, and this in fact agrees with the exact
definition of DFS. To see this, recall that the trace of a product of
matrices is invariant under cyclic permutation, and use Equations
\ref{eq:dfs-def}, \ref{eq:HAH-def} and \ref{eq:eigen-HAH} to derive
\begin{align}
 \mbox{DFS}= \tr \mathbf{HK}
 = \tr \mathbf{R}^{-1/2}\mathbf{HK}\mathbf{R}^{1/2}
 = \tr \mathbf{H'AH'}^T 
 = \tr \mathbf{U}\bm{\Lambda^a}\mathbf{U}^T 
 = \tr \bm{\Lambda^a}
 = \sum_{i=1}^r \lambda^a_i, \label{eq:dfs-trHAH}
\end{align}
where, in the penultimate equality, we have used the fact that
$\mathbf{U}^T\mathbf{U}$ is a unit matrix. A more detailed discussion along this line
can be found in \cite{Fisher2003}.

\subsection{Upper bounds of DFS}
Equation \ref{eq:dfs-trHAH} leads to obvious upper bounds of DFS. Recalling
that $\lambda^b_i ={\sigma^b_i}^2$ are all non-negative, it follows
immediately from Equation \ref{eq:eigval-A} that
\begin{align}
 0 \leq \lambda^a_i < 1,
\end{align}
from which follows that
\begin{align}
 \mbox{DFS}=\sum_{i=1}^r \lambda^a_i < r 
 = \mbox{rank\ }\mathbf{R}^{-1/2}\mathbf{HB}^{1/2}
 = \min\left\{\mbox{rank\ }\mathbf{R}, \mbox{rank\ }\mathbf{H}, \mbox{rank\ }\mathbf{B} \right\} \label{eq:dfs-bounds-1}\\
 = \min\left\{ \mbox{number of observations, dimension of state space}\right\}. \label{eq:dfs-bounds-2}
\end{align}
These upper bounds may seem self-evident as long as we deal with an
optimal DA scheme where both the observation and background error
covariance matrices, $\mathbf{R}$ and $\mathbf{B}$, are perfectly
prescribed (\ie identical to the true ones) and all the matrix
operations are performed exactly. However, as we will show in the next
section, these simple upper bounds become relevant and can provide
meaningful insight when we analyze practical algorithms which compromise
optimality for affordable computational complexity.

\section{DFS applied to ETKF} 
\label{sec:dfs-ens} In this section, 

we apply the theory of DFS just outlined above to ETKF
\citep{Bishop2001,Wang2004}, a determistic variant of EnKF, and show
that the DFS that is attained with this scheme can never exceed the
ensemble size. The same discussion also applies to each local analysis
of LETKF \citep{Hunt2007}, a local variant of ETKF. This seemingly
simple fact, the authors believe, has many important implications, as we
discuss later. In this paper we focus on ETKF and their variants, but
the results should be valid for other implementations of square-root
filters like Ensemble Adjustment Kalman Filter
\citep[EAKF][]{Anderson2001} and the serial ensemble square-root filter
\citep[EnSRF][]{WhitakerHamill2002} since these algorithms, when
performed without localization, have been shown to result in the same
posterior mean and the same error space spanned by the posterior perturbations
\citep{Tippett2003}.

\subsection{Proof of DFS being less than the ensemble size}
\label{sec:dfs-proof}
We consider the ETKF algorithm, or a local assimilation step of LEKTF,
and for now ignore covariance localization. In ETKF or LETKF, as with
any EnKF algorithms, the background error covariance matrix $\mathbf{B}$
is approximated using $K$ members of ensemble forecasts,
$\mathbf{x}^b_i,\ i=1,2,\cdots,K$, as
\begin{align}
 \mathbf{B}^\mathrm{ens} = \frac{1}{K-1}\mathbf{X}^b\mathbf{X}^{bT} \label{eq:def-Bens}
\end{align}
where $K$ is the ensemble size and the matrix $\mathbf{X}^b$
is the matrix of background perturbations defined as
 $\mathbf{X}^b=\left[\mathbf{\delta x}^b_1,\mathbf{\delta x}^b_2\cdots,\mathbf{\delta x}^b_K \right]$
where $ \mathbf{\delta x}^b_i=\mathbf{x}^b_i - \overline{\mathbf{x}}^b,
i=1,\cdots,K$ are the $N$-dimensional vectors of background
perturbations where $N$ is the dimension of the state space, and
$\overline{\mathbf{x}}^b=(1/K)\sum_{i=1}^K \mathbf{x}^b_i$ is the
ensemble mean of the background state. The discussion given in section
\ref{sec:dfs} remains valid for ETKF or each local analysis of LETKF (in
fact, these algorithms rely on Equation \ref{eq:svd} in computing
$\mathbf{A}$ and $\mathbf{K}$), so that Equation \ref{eq:dfs-bounds-1}
implies that the DFS for these algorithms (denoted
$\mbox{DFS}^\mathrm{ens}$ hereafter) is bounded from above by $K-1$:
\begin{align}
 \mbox{DFS}^\mathrm{ens} < K-1 \label{eq:dfs-ens-upperbounds}
\end{align}
since $\mbox{rank\ }\mathbf{B}^\mathrm{ens}=\mbox{rank\
}\mathbf{X}^b\leq K-1$.  If we define the optimal DFS (denoted
$\mbox{DFS}^\mathrm{opt}$ hereafter) as the DFS that would be attained
if the analysis is performed optimally with the canonical KF, the DFS
for ETKF (or local analysis of LEKTF) is unavoidably underestimated
provided that $\mbox{DFS}^\mathrm{opt}> K-1$.

In a practical NWP setup, the underestimation of DFS is quite likely;
for example, in the global LETKF of JMA (see the Appendix), the ensemble
size $K$ is only 50, while the number of observations that are
assimilated locally is typically $O(10^3)$ or even greater, so that
$\mbox{DFS}^\mathrm{opt}$ locally should be $O(10^2)$ or more
\footnote{Here we assume that the observations are of comparable
accuracy to the background. In such a case, the observations and the
background are roughly equally informative, so it should be legitimate
to assume, from Equation \ref{eq:self-sensitivity}, that the per-obs DFS
is, on average, not too different from one half, which entails that
$\mbox{DFS}^\mathrm{opt}$ should be of the same order of magnitude to
that of half the number of observations. }
 which is much larger than $K$. The implications of this DFS
underestimation are discussed in more detail below.

\subsection{Implications of DFS underestimation}
\label{sec:implications-dfs}

The underestimation of DFS (\ie, $\mbox{DFS}^\mathrm{ens} <
\mbox{DFS}^\mathrm{opt}$) means, by definition, that the analysis is not
fully extracting information from observations. More specifically, since
the analysis increment projected onto the observation space is
$\mathbf{HKd}$ where $\mathbf{d}$ is the O-B departure
$\mathbf{y}^o-\mathbf{H}\overline{\mathbf{x}}^b$, an
underestimated $\mbox{DFS}^\mathrm{ens}=\tr \mathbf{HK}^\mathrm{ens}$
(where $\mathbf{K}^\mathrm{ens}$ denotes the Kalman gain used in EnKF)
suggests that the analysis increment (or the correction of the background
by the observations) is likely smaller than what it should be under optimality.

A more important consequence of DFS underestimation is that the analysis
becomes overconfident (or equivalently, the ensemble becomes
underdispersive). This is becuase DFS coincides with (the square of) the
analysis spread measured in the normalized observation space since
$\mbox{DFS}=\tr \mathbf{H'AH'}^T \left(= \tr
\mathbf{R}^{-1/2}\mathbf{HAH}^T\mathbf{R}^{-1/2} \right)$ (see Equation
\ref{eq:dfs-trHAH}). Consider a situation where we assimilate $p$
independent observations, each of which is as accurate as their
counterpart from the background, in which case the DFS that should be
attained under optimality should be of comparable order of magnitude to
the number of the assimilated observations: $\mbox{DFS}^\mathrm{opt}=c
\times p$ with some positive $c \sim O(0.1)$ (see the footnote in
section \ref{sec:dfs-proof} for a rationale behind this rough
estimate). Note here that the DFS attained under optimality, or the
posterior error variance (measured in the normalized observation space)
that the optimal analysis scheme assumes, is by definition identical to
the statistical mean of its true value:
\begin{align}
 \mbox{DFS}^\mathrm{opt} = \mathrm{E}\left\|\mathbf{H'}\left(\mathbf{x}^\mathrm{a,opt}-\mathbf{x}^\mathrm{true}\right)\right\|^2
\end{align}
where $\mathrm{E}(\cdot)$ denotes statistical expectation,
$\mathbf{x}^\mathrm{a,opt}$ denotes the posterior state obtained by an
optimal analysis, and $\mathbf{x}^\mathrm{true}$ denotes the true state.
Now, if we assimilate such observations by an ETKF with the ensemble
size $K$ that is orders of magnitude smaller than the number of
observations $p$ (\ie, $K \ll c\times p=\mbox{DFS}^\mathrm{opt}$), it
follows from Equation \ref{eq:dfs-ens-upperbounds} that
\begin{align}
\tr \mathbf{H'A}^\mathrm{ens}\mathbf{H'}^T
 = \mbox{DFS}^\mathrm{ens} \ll \mbox{DFS}^\mathrm{opt} 
 = \mathrm{E}\left\|\mathbf{H'}\left(\mathbf{x}^\mathrm{a,opt}-\mathbf{x}^\mathrm{true}\right)\right\|^2
 < \mathrm{E}\left\|\mathbf{H'}\left(\overline{\mathbf{x}}^\mathrm{a,ens}-\mathbf{x}^\mathrm{true}\right)\right\|^2 \label{eq:ens-underdispersion}
\end{align}
where $\overline{\mathbf{x}}^\mathrm{a,ens}$ denotes the ensemble mean
of the posterior state, and the rightmost inequality follows from an
assumption that ETKF with a limited ensemble size should result in an
analysis inferior to that of the optimal KF. Equation
\ref{eq:ens-underdispersion} means that the squared posterior spread,
which is the estimated posterior error variance assumed by the ETKF, is
much smaller than the true error variance of the posterior mean (or
equivalently, the posterior ensemble is underdispersive).

The overconfidence of analysis (or the uderdispersion of the posterior
ensemble) can accumulate over cycles and can eventually lead to filter
divergence. Various covariance inflation techniques have been proposed
and are employed to counteract against it. The discussion given in the
paragraph above suggests that very strong covariance inflation is
required if the ensemble size $K$ is much smaller than
$\mbox{DFS}^\mathrm{opt}$, but covariance inflation that is too strong
is undesirable because that would ruin the EnKF's ability to represent
the ``errors of the day,'' which is one of the most appealing aspects of
EnKF algorithms.

We can also infer that a small ensemble size $K$ can be tolerated if
$\mbox{DFS}^\mathrm{opt}$ is small. Such a situation can happen, for
example, when
\begin{itemize}
 \item observations are much less accurate than the background (so that
       all singular values of the true observability matrix
       ($\mathbf{H}'{\mathbf{B}^\mathrm{true}}^{1/2}$) become much
       smaller than one), or
 \item the singular spectrum of the true observability matrix is dominated
       by a small number of large ones, which occurs if the dynamical
       system has a small number of growing modes so that the error space
       (locally) has low unstable dimensions.
\end{itemize}
Conversely, given an ensemble size $K$, we can avoid the underdispersion
by limiting the number of observations to assimilate to $K$ times some
factor of $O(1)$ by, say, some form of thinning or by choosing a tighter
domain localization (in the case of LETKF) for areas that abound with
observations. Such measures come at the price of discarding some pieces
of information from observations, but are nevertheless shown by a number
of previous studies to be very effective in practice, as we discuss in
section \ref{sec:discussion-less-is-good}.

\subsection{Role of localization}
\label{sec:role-of-loc}
In the previous subsections, we deliberately deferred discussing the
impact of covariance localization on DFS; this subsection is devoted to
exploring this issue.

In the context of EnKF, covariance localization has traditionally been
explained as serving two different (but related) functions: one is to
suppress spurious correlations that appear in covariance matrices
because of sampling noises, and the other is to mitigate the so-called
``rank issue'' (or ``rank deficiency issue'') that is roughly defined as
any issues that arise from the ensemble-derived background covariance
$\mathbf{B}^\mathrm{ens}$ not being full-rank
\cite[e.g.][]{HoutekamerZhang2016}. Arguments based on DFS concept helps
us to quantitatively assess the latter (\ie, how localization mitigate
the rank issue).

Covariance localization schemes generally applied in EnKF algorithms can
be classified into two types depending on whether they operate on the
observation error covariance matrix $\mathbf{R}$ or on the background
error covariance matrix $\mathbf{B}$ \citep{Greybush2011}. The former
(called ``R-localization'' hereafter) is typically used with LETKF. The
latter (called B-localization hereafter) can be further split into
``observation-space B-localization'' that operates on $\mathbf{BH}^T$
and $\mathbf{HBH}^T$ in computing the gain $\mathbf{K}$, and
``model-space B-localization'' that operates directly on
$\mathbf{B}$. In this paper we focus on the difference between
R-localization and model-space B-localization.

R-localization implements localization by inflating the error variance
for observations that are far from the analyzed grid point. This can be
realized by replacing $\mathbf{R}$ with $\mathbf{R}_\mathrm{loc} =
\mathbf{L}_g\mathbf{R}\mathbf{L}_g$ in each local analysis where
$\mathbf{L}_g$ is a $p\times p$ diagonal matrix whose diagonal entries
are the values of some increasing function (typically the inverse
square-root of the Gaussian function) of the distance between the
analyzed grid point and the location of the observation. Simply
replacing $\mathbf{R}$ with $\mathbf{R}_\mathrm{loc}$ has no effect on
the upper bound on $\mbox{DFS}^\mathrm{ens}$ given in Equation
\ref{eq:dfs-ens-upperbounds}, suggesting that R-localization, while
effective in suppressing spurious correlations, does not help in
mitigating the rank issue. In section \ref{sec:exp}, we show that
R-localization may even decrease DFS, which is also demonstrated in
\cite{Huang2019}. While R-localization (artificial inflation of the error
variance for distant observations) in itself does not lend to mitigate
the rank issue, domain localization (c.f., section
\ref{sec:exp-role-of-R-loc}) that is inherent in R-localization can
mitigate the issue. We will discuss this issue further in section
\ref{sec:exp-role-of-R-loc}.

Model-space B-localization implements localization by tapering the
$\mathbf{B}$ matrix through taking Schur product (element-wise
multiplication) with a localization matrix $\bm{\rho}$ whose
$(i,j)$-element is the value of some decreasing function of the distance
between the locations of $i$-th and $j$-th elements of the state
vector. Implementing this type of localization in an EnKF is not
straightforward since $\mathbf{B}$ matrix is not explicitly constructed
in EnKF, but the ensemble modulation approach, proposed in
\cite{BishopHodyss2009b}, allows us to perform model-space B-localization without
explicitly constructing $\mathbf{B}$ in model-space. In this
approach, the ensemble covariance localized with $\bm{\rho}$ is
expressed as a sample covariance of a larger ensemble:
\begin{align}
\bm{\rho}\circ\mathbf{B}^\mathrm{ens}
 = \frac{1}{K-1}\bm{\rho}\circ \left(\mathbf{X}^b{\mathbf{X}}^T\right)
 = \frac{1}{M-1}\mathrm{ZZ}^T \label{eq:mod-prod}
\end{align}
where $M=KL$, $L$ is the rank of $\bm\rho$, and the $N\times M$
``modulated ensemble'' $\mathbf{Z}$ is defined, using the square-root
matrix $\mathbf{L}=\left[\mathbf{l}_1,\cdots,\mathbf{l}_L\right]$ of
$\bm\rho$ that satisfies $\mathbf{LL}^T=\bm{\rho}$, as
\begin{align}
 \mathbf{Z}=\sqrt{\frac{M-1}{K-1}}\left[
 \left(\mathbf{l}_1\circ\mathbf{\delta x}^b_1, \mathbf{l}_2\circ\mathbf{\delta x}^b_1,\cdots,\mathbf{l}_L\circ\mathbf{\delta x}^b_1\right),\cdots,
 \left(\mathbf{l}_1\circ\mathbf{\delta x}^b_K, \mathbf{l}_2\circ\mathbf{\delta x}^b_K,\cdots,\mathbf{l}_L\circ\mathbf{\delta x}^b_K\right)
 \right].
\end{align}
From Equation \ref{eq:mod-prod} we can see that performing a regular
EnKF algorithm using $\mathbf{Z}$ as the ensemble of background
perturbations in place of $\mathbf{X}^b$ achieves model-space
B-localization. We remark that choosing $\mathbf{L}$ as the exact square
root of $\bm\rho$ results in $L$ being the rank of $\bm\rho$, which is
unaffordably large, so in practice, $\mathbf{L}$ is approximated by
retaining only the dominant eigen modes of $\bm\rho$. For ETKF, this
model-space B-localization through ensemble modulation has difficulty in
updating perturbations since it results in $M=KL$ posterior members
produced given $K$ prior members, but we need $K$-member ensemble to
initialize the next cycle. A method that resolves this difficulty have
recently been devised independently by \cite{Bocquet2016} and
\cite{BWL17}.

By model-space B-localization through ensemble modulation, the
upper bound on DFS given in Equation \ref{eq:dfs-ens-upperbounds}
increases $L$-fold from $K-1$ to $KL-1$, suggesting that DFS
underestimation (and hence underdispersion of posterior ensemble) can be
potentially mitigated. In section \ref{sec:exp}, we demonstate, with a
simple covariance model, that model-space B-localization does indeed
significantly increase $\mbox{DFS}^\mathrm{ens}$.

\subsection{Impact of covariance inflation on DFS}
It is worth mentioning that several covariance inflation methods act to
increase $\mbox{DFS}^\mathrm{ens}$ but their impact is limited since the
upper bounds given in Equation \ref{eq:dfs-ens-upperbounds} still
applies even after the introduction of such methods.  With the
multiplicative inflation \citep{Pham1998,AndersonAnderson1999}, each
prior perturbation is inflated by a common factor $a>1$, resulting in
each $\lambda^b_i$ uniformly inflated by the factor $a^2>1$; recalling
that $\lambda^a_i$ as a function of $\lambda^b_i$ is monotinically
increasing (see Equation \ref{eq:eigval-A}), this means that each
$\lambda^a_i (i=1,\cdots,r)$, and hence $\mbox{DFS}^\mathrm{ens}$ (which
is their sum over all $i$'s), are also increased. Similarly, with the
additive inflation \citep{MitchellHoutekamer2000}, a random independent
draw from a certain predefined error distribution is added to each prior
perturbation (each column the matrix $\mathbf{X}^b$) before performing
analysis, leading to $\mathbf{B}^\mathrm{ens}$ in Equation
\ref{eq:def-Bens} replaced by $\mathbf{B}^\mathrm{ens}$ plus some
symmetric matrix $\mathbf{Q}^\mathrm{ens}$. Since this
$\mathbf{Q}^\mathrm{ens}$ is a positive-semidefinite matrix of rank at
most $K-1$, all $\lambda^b_i (i=1,\cdots,r \ll K-1)$ are added with some
positive increment, resulting in $\mbox{DFS}^\mathrm{ens}$ increased
accordingly. However, despite being increased by these inflation
methods, $\mbox{DFS}^\mathrm{ens}$ is still subject to the upperbound
(Equation \ref{eq:dfs-ens-upperbounds}) since the rank of
$\mathbf{B}^\mathrm{ens}$ does not increase by these operations.

The inflation methods that operate on the posterior perturbations such
as RTPP and RTPS (see section \ref{sec:discussion-inflation}) are
typically applied after computing the gain matrix $\mathbf{K}$. As such,
while $\mbox{DFS}^\mathrm{ens}$ could be increased by these methods, the
underestimation of the analysis increment $\mathbf{Kd}$ cannot be mitigated.

\subsection{Comparison with previous literature}

The most important messages from the examination of DFS given above are
that, if the ensemble size $K$ is insufficient relative to
$\mbox{DFS}^\mathrm{opt}$ (the true information content of the
observations), posterior ensemble from EnKF algorithms like ETKF or
LETKF will be automatically underdispersive, which hinders effective
exploitation of observational information, and that this limitation can
be mitigated by model-space B-localization that applies Schur product
tapering on $\mathbf{B}$.  We remark that \cite{FurrerBengtsson2007}
obtained a similar result for stochastic EnKF with perturbed
observations: they showed, for special cases where
$\mathbf{HH}^T=\mathbf{I}_p$ holds, that, the concavity of $\mathbf{HK}$
as a function of the background error covariance (in the observation
space) $\mathbf{HBH}^T$ implies negatively-biased expectation of $\tr
\mathbf{HK}$ via Jensen's inequality, which, with our notation, can be
summarized as
\begin{align}
 \mathrm{E}\left[\mbox{DFS}^\mathrm{ens}\right] \leq \mbox{DFS}^\mathrm{opt}.
\end{align}
\cite{WhitakerHamill2002} also remarked this property for a univariate
case by experimentation. Compared to their findings, our proposition in
Equation \ref{eq:dfs-ens-upperbounds} is stronger in being valid
deterministically (not being valid only in expected sense), and in
giving a more explicit upper bound.

Using the interpretation of DFS given in the last paragraph of section
2.1, $\mbox{DFS}^\mathrm{ens} < K$ can be expressed in plain language as
``there are at most $K$ directions in which the background can be
adjusted to fit the observations.'' This is exactly what
\cite{Lorenc2003} pointed out in his section 3(b) through his informal
(but insightful) contemplation. \cite{Lorenc2003} also explained how
localization on $\mathbf{B}$ should possibly alleviate this limitation,
again consistent with our discussion given above, but did not explicitly
discuss how this is linked to underdispersive (or overconfident)
posterior covariance.

\section{Exposition with a simple one-dimensional covariance model} 
\label{sec:exp}

The argument developed in the previous section using the concept of DFS
gives us useful insights as to in what conditions analysis by (L)ETKF
becomes suboptimal. In this section we provide illustrative examples in a
conceivably simplest setup. We consider an idealistic situation where (1) the
assimilations are not cycled, (2) the true error covariances
$\mathbf{B}$ and $\mathbf{R}$ are known, and (3) the background
perturbations $\mathbf{X}^b$ are generated perfectly (in the sense that
$\frac{1}{K-1}\mathbf{X}^b{\mathbf{X}^b}^T$ converges to the true
$\mathbf{B}$ as $K\rightarrow \infty$). The detailed set-up is given
below, followed by descriptions of some illustrative results.

\subsection{Experimental set-up}
We consider a state space that results from discretizing a periodic
one-dimensional domain into equi-spaced $N_\mathrm{grid}=360$ grid
points, so that the state can be represented by a
$N_\mathrm{grid}$-dimensional column vector $\mathbf{x}$. 

\begin{figure}[htbp]
 \centering
 \includegraphics[bb=7 6 590 398, width=0.45\textwidth]{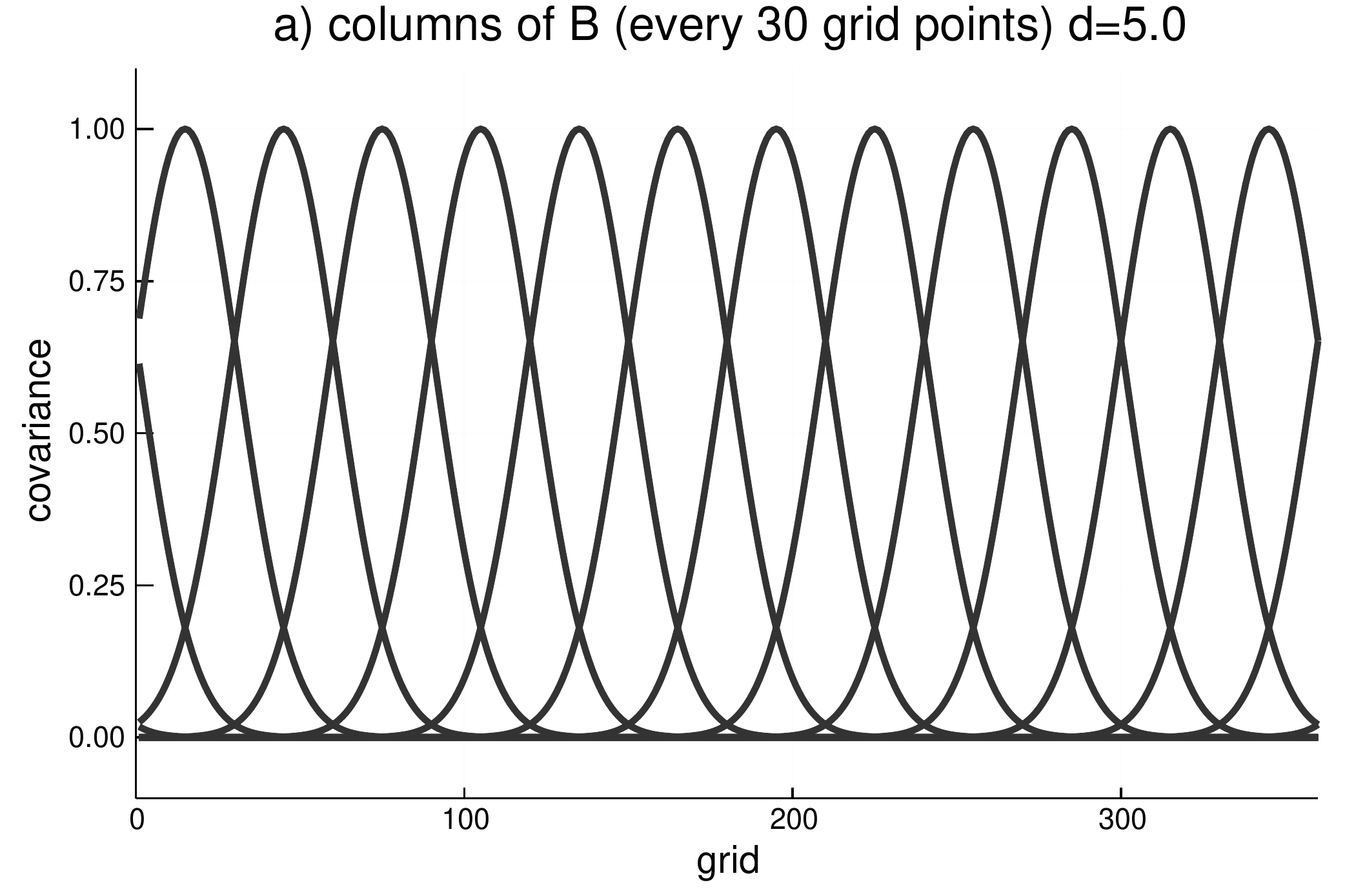} 
 \includegraphics[bb=7 6 590 398, width=0.45\textwidth]{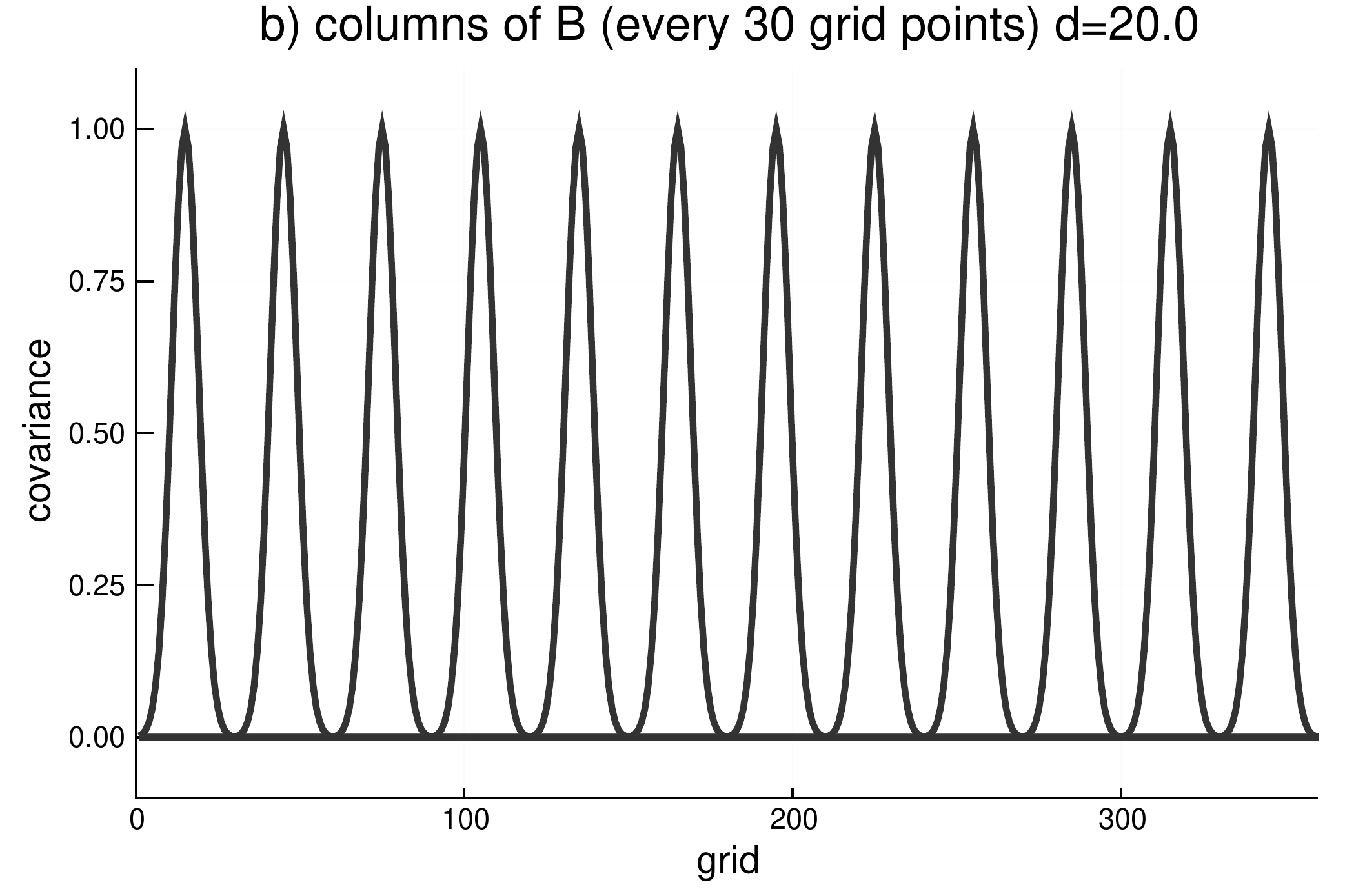} 
 \caption{
 Structure of the true background error covariance matrix $\mathbf{B}$ for
 parameters (a) $d=5$ and (b) $d=20$. Each curve represents the column
 of $\mathbf{B}$ starting from the 15th column with a stride of
 30. Larger values of $d$ correspond to narrower peaks.
 } \label{fig:B-structure}
\end{figure}

As the background error covariance model, we follow the sinusoidal basis
model given in the Appendix A.1 of \cite{BishopHodyss2009a}. In this
model, the covariance matrix is diagonalized in the space spanned by the
sinusoids (sine and cosine curves) with wavenumbers at most
$N_\mathrm{grid}/2$, and the eigenvalues are chosen such that each
column (or raw) of the covariance matrix takes an identical Gaussian
shape whose length scale is controlled by a parameter $d$. The variances
at each grid point (the diagonal entries of the covariance matrix) are
all set to one.  Visual depiction is perhaps more helpful than the
precise definition given with equations; instances of this covariance
model, with parameter $d=5$ and 20, are shown in Figure
\ref{fig:B-structure}. All the experiments shown below are performed
with $d=20$ unless otherwise stated, but we have also repeated all the
experiments with different choices of $d$ and confirmed that the results
are qualitatively similar.

We assume that the state is observed at every 3 grid points including
the first grid point indexed with 1, resulting in $p=120$ observations
which can be represented by a $p$-dimensional column vector
$\mathbf{y}^o$. We assume that the observation errors are uncorrelated
and each observation share the same error variance ${\sigma^o}^2$, so
that $\mathbf{R}={\sigma^o}^2\mathrm{I}_p$.

With this set-up, we conducted ETKF or LETKF assimilation experiments to
investigate their properties in light of DFS argument. As a reference, we
also conducted canonical KF analysis using the true error
covariances. For each of the ETKF experiments, we stochastically
generate the observations, the background mean and the background
perturbations, respectively, by:
\begin{align}
 \mathbf{y}^o &= \mathbf{H}\mathbf{x}^\mathrm{true} + \mathbf{R}^{1/2}\bm{\varepsilon}^o\\
 \mathbf{x}^b &= \mathbf{x}^\mathrm{true} +  {\mathbf{B}^\mathrm{true}}^{1/2}\bm{\varepsilon}^b \\
 \mathbf{X}^b &= {\mathbf{B}^\mathrm{true}}^{1/2}\left[\bm{\varepsilon}_1,\bm{\varepsilon}_2,\cdots,\bm{\varepsilon}_K\right] 
\end{align}
where $\mathbf{H}$ is the observation operator which picks up the values
of the state vector $\mathbf{x}$ at every 3 grid points starting from
the first (topmost) element of $\mathbf{x}$, $\mathbf{x}^\mathrm{true}$
is the hypothetical true state, ${\mathbf{B}^\mathrm{true}}^{1/2}$ is
the symmetric square root of the true background error covariance matrix
${\mathbf{B}^\mathrm{true}}$ constructed as stated above, $K$ is the
ensemble size, and the vectors $\bm{\varepsilon}^o \in \mathbb{R}^p$,
$\bm{\varepsilon}^b \in \mathbb{R}^{N_\mathrm{grid}}$ and
$\bm{\varepsilon}_i \in \mathbb{R}^{N_\mathrm{grid}} (i=1,\cdots,K)$ are
random vectors, each entry of which is an independent draw from the
standard normal distribution with mean zero and unit variance. The
choice of the true state $\mathbf{x}^\mathrm{true}$ is irrelevant in
this study, so we arbitrarily set it to a zero vector.

To highlight the relevance of DFS in terms of the ability of DA methods
to extract observational information, we consider two different
scenarios. In the first, ``high DFS'' scenario, we set $\sigma^o=1$. In
this case, the observations are about as accurate as the background
(recall that we chose $\mathbf{B}^\mathrm{true}$ so that each of its
diagonal entries are one), meaning that observations should provide
about the same amount of information as the background does. We have
equally accurate sources of information from 360 background values and
120 observation, so that we can intuitively expect the optimal DFS to be
close to $120/(120+360)$ times the number of observations (120), which
is 30, and this is indeed not very different from the exact value of the
optimal DFS which is 39.877. In the second, ``low DFS'' scenario, we set
$\sigma^o=5$, so that observations are five times less accurate than the
corresponding background. In this case, observations provide much less
information compared to the background, so that the optimal DFS should
be small. The exact value of the optimal DFS in this case is 4.386.

For each of the senario, we conduct ETKF experiments, with or without
localization, using different ensemble sizes. In each experiment we
conduct 1,000 trials changing the seed for the random number generator
to ensure statistical robustness of the results.

In experiments that apply covariance localization, we use the 5th order
piecewise rational function with finite support defined in Eq. (4.10) of
\cite{GaspariCohn1999}. This function becomes identically zero beyond a
cut-off distance $d_\mathrm{cut-off}$, and hereafter we use this cut-off
distance to specify the length scale of localization.

 In the experiments discussed in section \ref{sec:exp-role-of-loc}, we
implemented B-localization as model-space B-localization through the
ensemble modulation technique. We remark that, with our observation
operator $\mathbf{H}$ that only picks up state variables at selected
grid points, model-space and observation-space B-localization are
equivalent.

\subsection{Dependence on the ensemble size}

\begin{figure}[htbp]
 \centering
 \includegraphics[bb=0 0 590 400, width=0.45\textwidth]{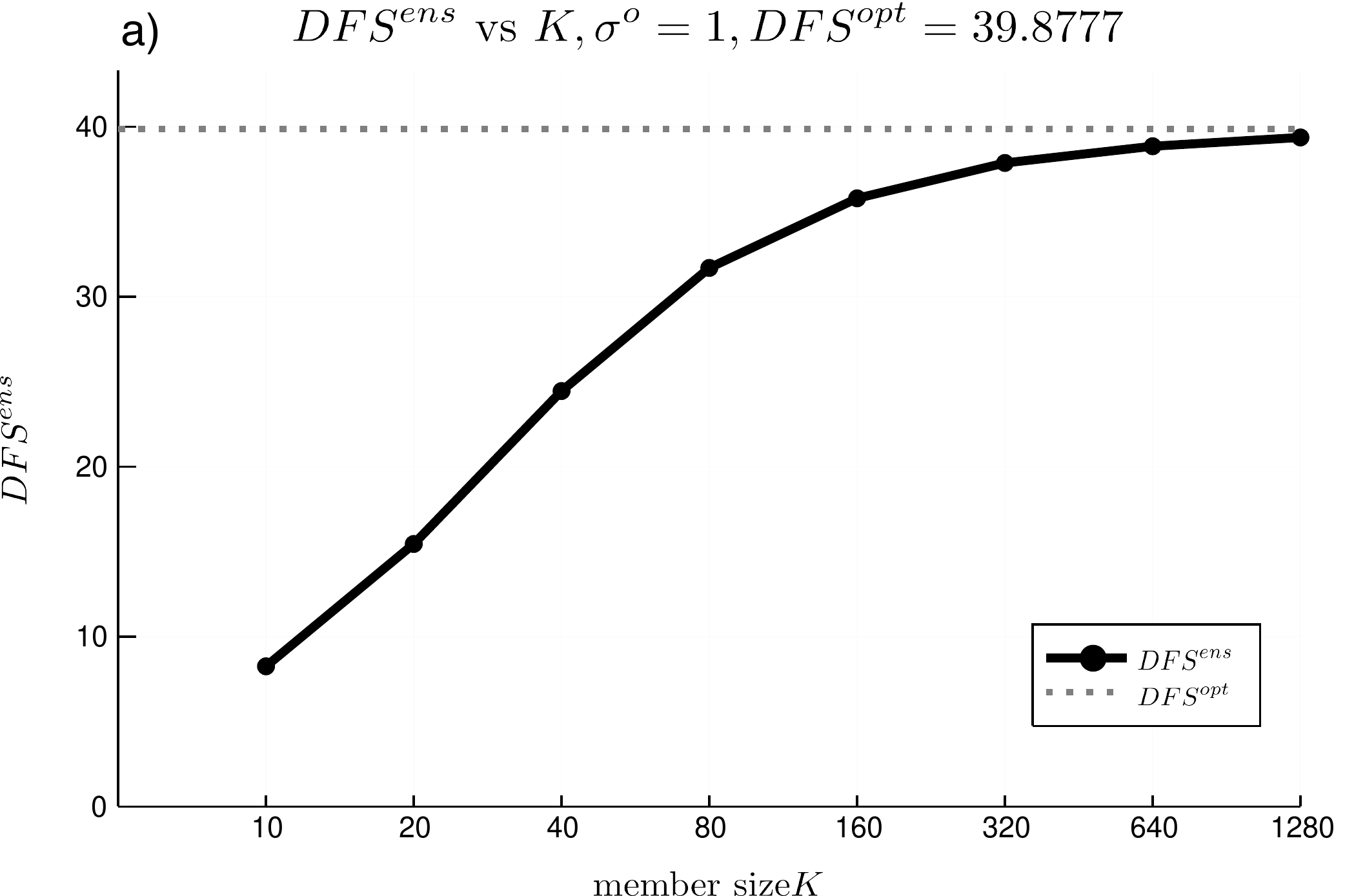}  
 \includegraphics[bb=0 0 590 400, width=0.45\textwidth]{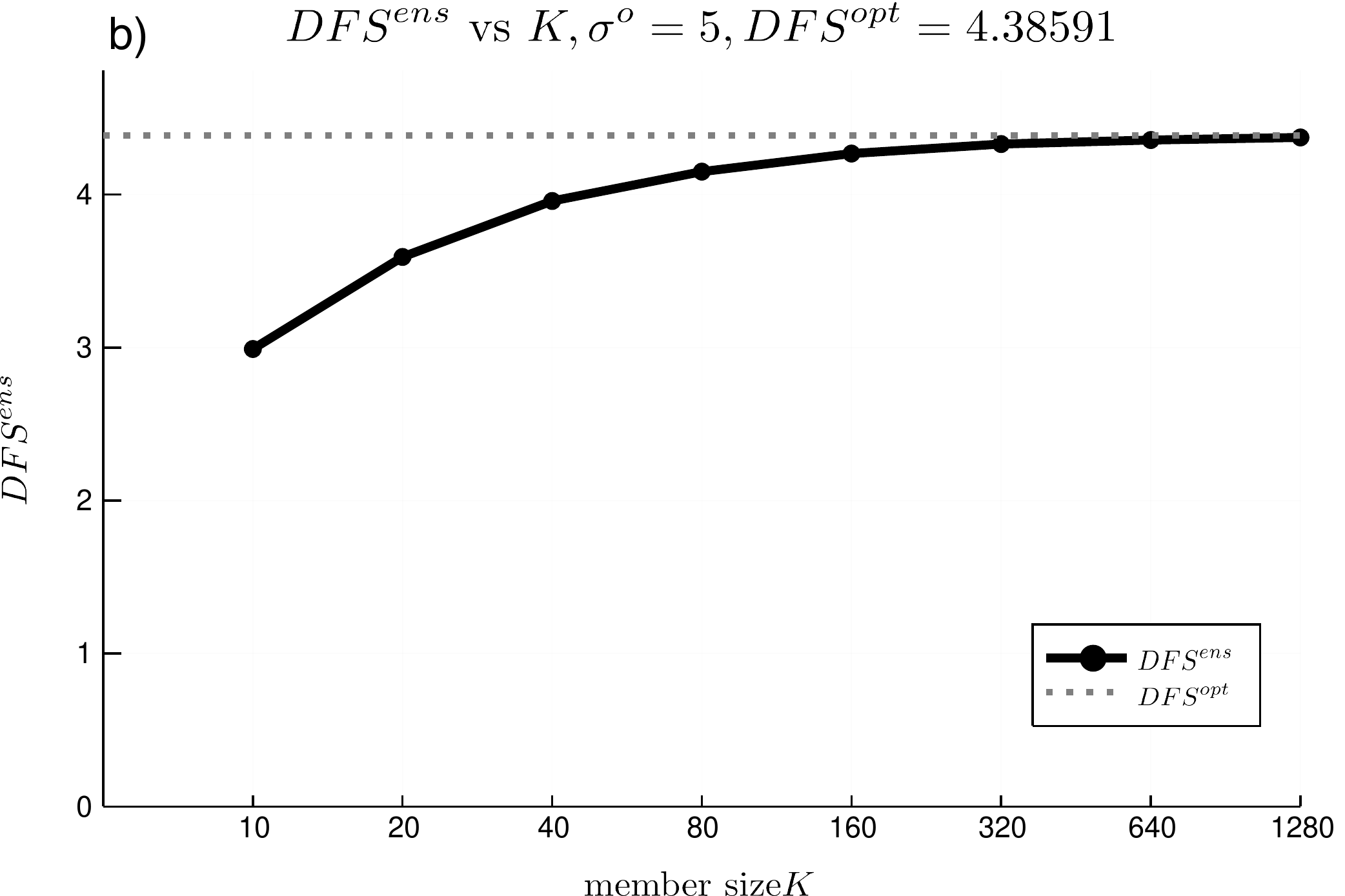}    
 \caption{
 DFS attained by ETKF (or each local analysis of LETKF),
 $\mbox{DFS}^\mathrm{ens}$, plotted as a function of the ensemble size
 $K$, for (a) the ``high DFS'' and (b) ``low DFS'' scenarios. As a
 reference, the DFS that would be attained by an optimal canonical KF,
 $\mbox{DFS}^\mathrm{opt}$, is plotted in each panel as a horizontal
 dashed line.
 } \label{fig:dfs-vs-K}
\end{figure}

We first examine cases without localization. The dependence of
$\mbox{DFS}^\mathrm{ens}$ (averaged over 1,000 independent trials) on
the ensemble size $K$ is depicted in Figure \ref{fig:dfs-vs-K} together
with $\mbox{DFS}^\mathrm{opt}$ that should be attained by the optimal
canonical KF.  In both ``high DFS'' (panel a) and ``low DFS'' (panel b)
scenarios, $\mbox{DFS}^\mathrm{ens}$ monotonically increases as the
ensemble size $K$ gets larger until it converges to
$\mbox{DFS}^\mathrm{opt}$. In the ``high DFS'' scenario, approximately
200 members are required to achieve 90 \% of $\mbox{DFS}^\mathrm{opt}$,
whereas, in the ``low DFS'' scenario, the same level of saturation is
achieved with only $\sim$50 members. This contrast is consistent with our
expectation from the theory that a small ensemble size $K$ should be
tolerated if $\mbox{DFS}^\mathrm{opt}$ is small (c.f., section
\ref{sec:implications-dfs}).


The DFS diagnostics shown above is performed in the normalized
observation space, but similar tendencies can also be observed in the model
space. Figure \ref{fig:trA-mse-vs-K} shows the trace of the analysis
ensemble covariance, $\tr \mathbf{A}^\mathrm{ens}$, and the mean squared
error (MSE) of the analysis mean,
$\|\overline{\mathbf{x}}^\mathrm{a,ens} - \mathbf{x}^\mathrm{true}\|^2$,
averaged over 1,000 trials, as a function of the ensemble size $K$. As a
reference, their counterpart in an optimal canonical KF is also plotted with
a dotted line. In both ``high DFS'' and ``low DFS'' scenarios, both $\tr
\mathbf{A}^\mathrm{ens}$ and the analysis mean MSE converge to their
optimal value as $K$ becomes large, but the former is consistently
smaller than the latter. Recalling that $\tr \mathbf{A}^\mathrm{ens}$ is
the estimate of the analysis mean MSE that is assumed by the
assimilation algorithm, $\tr \mathbf{A}^\mathrm{ens}$ being smaller than
the analysis MSE means that the analysis is overconfident. We can
observe that, in both cases, the level of overconfidence diminishes as
the ensemble size $K$ gets larger. Comparing the two scenarios, we can
also observe that the overconfidence is much stronger in the ``high
DFS'' scenario than in the ``low DFS'' scenario. These results
corroborate our deduction from the theory that the ensemble size
required to alleviate overconfidence in analysis should increase in
proportion to $\mbox{DFS}^\mathrm{opt}$.

As discussed in section \ref{sec:implications-dfs}, underestimation of
DFS is suggestive of underestimation of analysis increment. A plot (not
shown) comparing the $l_2$-norm of analysis increment from ETKF,
$\|\mathbf{K}^\mathrm{ens}\mathbf{d}\|$ as a function of the ensemble
size $K$, with that of an optimal analysis,
$\|\mathbf{K}^\mathrm{opt}\mathbf{d}\|$, exhibits a converging curve
similar to the one shown in Figure \ref{fig:dfs-vs-K}, with the former
consistently underestimating the latter, again corroborating the
expectation from the theory.

\begin{figure}[htbp]
 \centering
 \includegraphics[bb=0 0 590 400, width=0.45\textwidth]{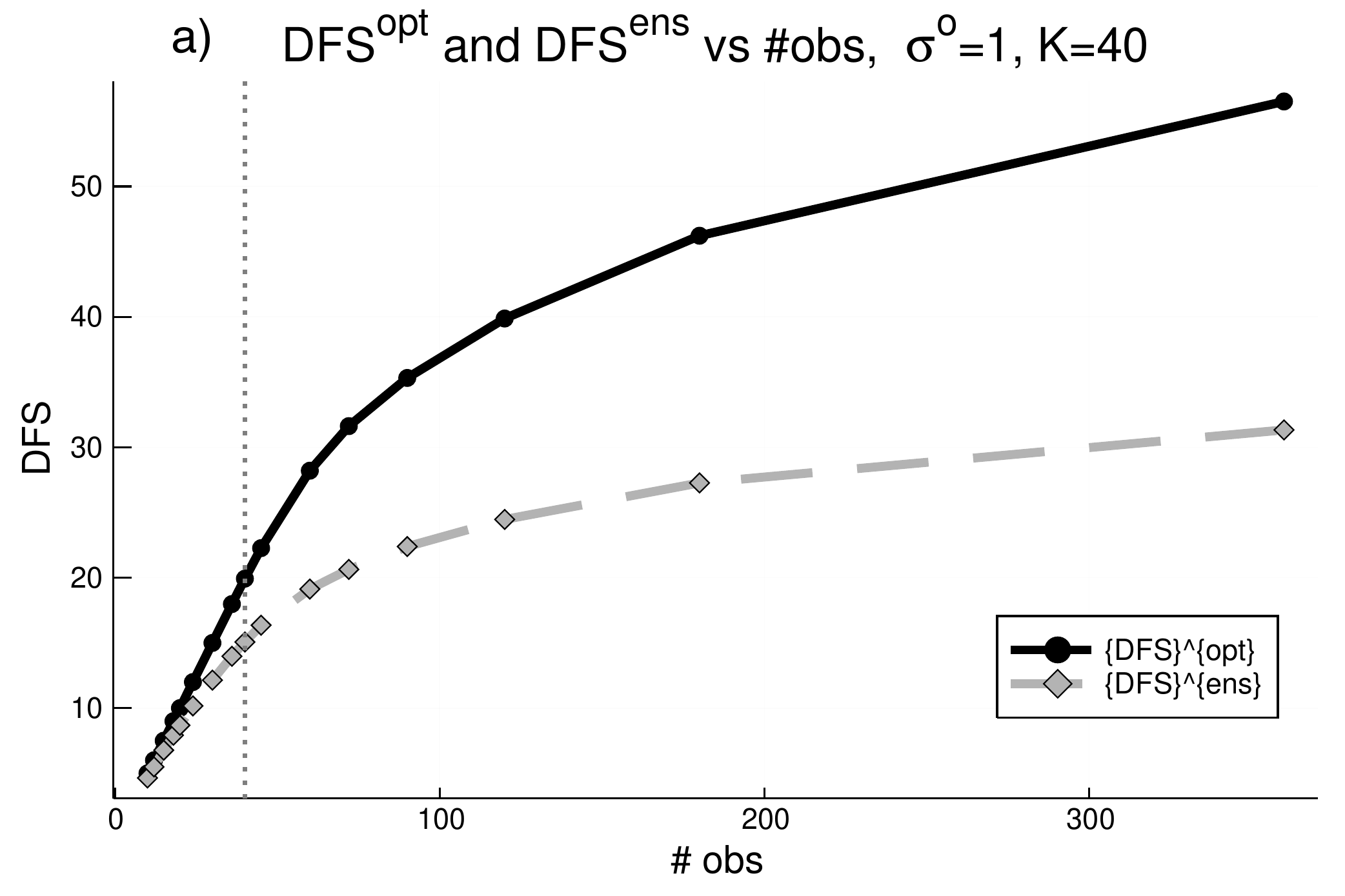}  
 \includegraphics[bb=0 0 590 400, width=0.45\textwidth]{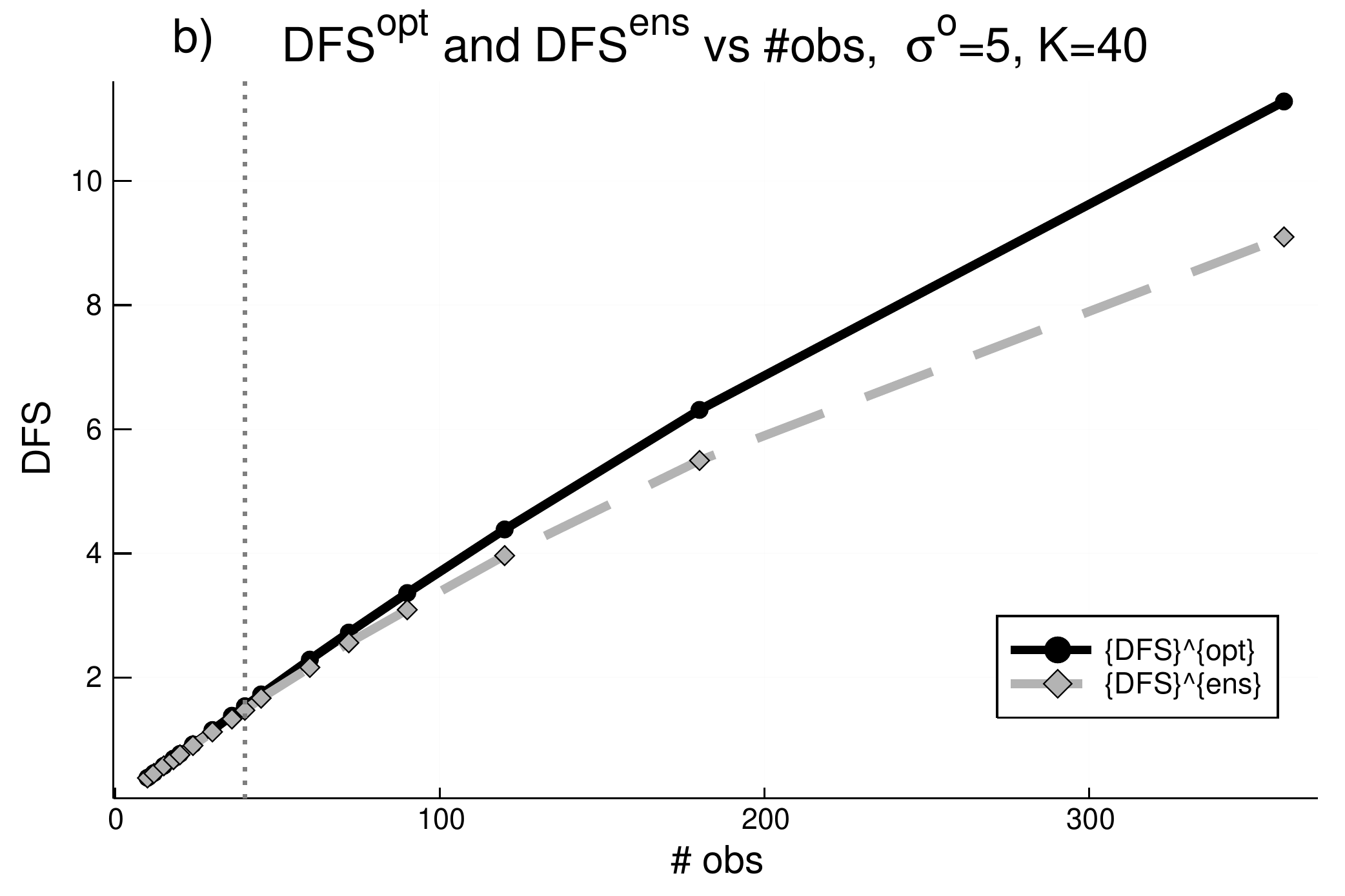}    
 \caption{
 The DFS attained by ETKF with the ensemble size $K=40$
 ($\mbox{DFS}^\mathrm{ens}$, gray dashed line) and the DFS attained by
 the optimal canonical KF ($\mbox{DFS}^\mathrm{opt}$, black solid line)
 plotted as a function of the number of observations, for (a) the ``high
 DFS'' and (b) ``low DFS'' scenarios. The vertical line in each panel
 shows the ensemble size $K=40$.
 } \label{fig:dfs-vs-numobs}
\end{figure}

A plot similar to Figure \ref{fig:dfs-vs-K} but with a fixed ensemble
size ($K=40$) and varying the number of observations (Figure
\ref{fig:dfs-vs-numobs}) is also illuminating. In the ``high DFS''
scenario, $\mbox{DFS}^\mathrm{ens}$ is close to
$\mbox{DFS}^\mathrm{opt}$ when the number of observations $p$ is small,
but as $p$ increase beyond the ensemble size $K=40$, the former begins
to underestimate the latter. In contrast, in the ``low DFS'' scenario,
$\mbox{DFS}^\mathrm{ens}$ stays close to $\mbox{DFS}^\mathrm{opt}$ even
for very large values of $p$. This is because $\mbox{DFS}^\mathrm{opt}$
is well below the ensemble size ($K=40$) even for the fully observed
case so that the $\mbox{DFS}^\mathrm{ens}$ being bounded by $K$ does not
pose much limitation.

\subsection{Role of localization}
\label{sec:exp-role-of-loc}
In section \ref{sec:role-of-loc} we expounded on how the DFS
underestimation that occurs if the ensemble size $K$ is much smaller
than $\mbox{DFS}^\mathrm{opt}$ can be alleviated by localization. In
this subsection we experimentally illustrate how the two localization
methods differ in this respect.

\subsubsection{B-localization}
Recalling that DFS is the sum of all the eigenvalues $\lambda^a_i$ of
the matrix $\mathbf{H'AH'}^T$, it is illuminating to examine how
localization changes the eigenspectrum of this matrix. 

\begin{figure}[htbp]
 \centering
 \includegraphics[bb=0 0 590 400, width=0.45\textwidth]{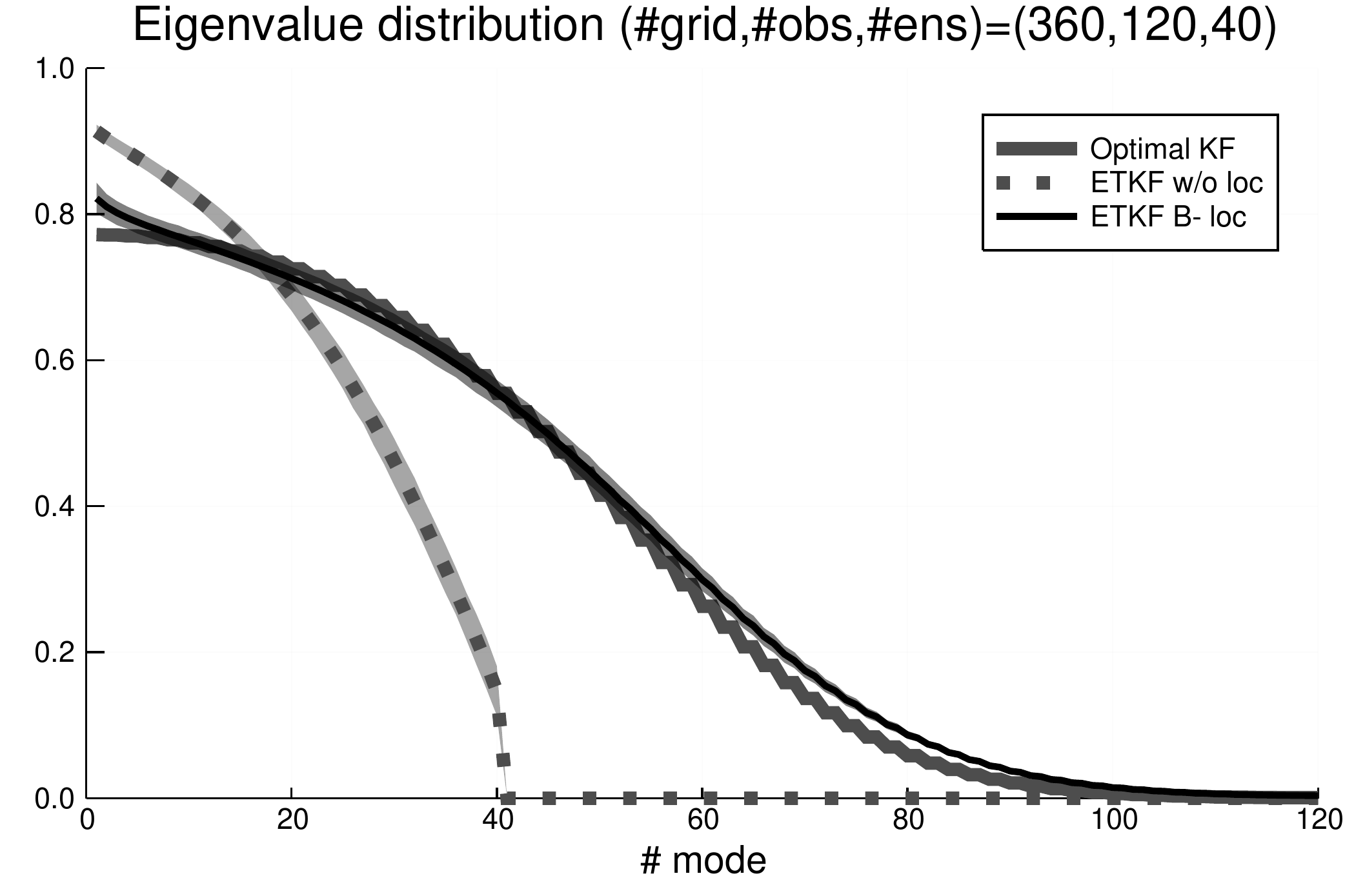} 
 \caption{
 Eigenvalues of the matrix $\mathbf{H'A}^\mathrm{opt}\mathbf{H'}^T$
 (gray thick solid line), $\mathbf{H'A}^\mathrm{ens}\mathbf{H'}^T$ (gray
 thick dotted line), and their counterpart for ETKF with model-space
 B-localization (black thin solid line) computed using the modulated
 ensemble. Shown by the lines are the mean over 1,000 trials. Their
 sampling variability, defined here as the upper and lower 5
 percentiles, is represented by the shades.
 } \label{fig:eig-HAH}
\end{figure}

The eigenvalues, sorted from the largest to the smallest, of the matrix
$\mathbf{H'A}^\mathrm{opt}\mathbf{H'}^T$ computed using the true
background covariance matrix with the canonical KF, are plotted in
Figure \ref{fig:eig-HAH} with the thick solid line. The true posterior
eigenvalues smoothly decrease as the mode number gets higher and almost
(but not exactly) vanishes at $\sim$ 100th and higher modes. Their
ensemble equivalent, computed by raw ETKF without any localization using
$\mathbf{B}^\mathrm{ens}$ constructed from 40 members (the thick dotted
line), abruptly become zero at the 40th mode. This is an indication of
DFS underestimation since the areas below these curves correspond to
DFS.

Model space B-localization allows us to avoid this abrupt truncation of
the eigenspectrum. The posterior eigenvalues, computed by ETKF
using the modulated background ensemble (see section
\ref{sec:role-of-loc}), are plotted with the thin solid line in Figure
\ref{fig:eig-HAH}. Here, we manually tuned the localization scale to 
achieve minimal analysis mean MSE (giving
$d_\mathrm{cut-off}=20\sqrt{10/3}\approx 36$),
%
%
and the localization matrix $\bm\rho$ is approximated by retaining the
leading $L=20$ eigen modes, which recovers 93.4\% of the trace of the
original matrix $\bm\rho$. We can observe that  a well-tuned B-localization
can almost perfectly recover the true posterior eigenspectrum, which
means that DFS underestimation can be avoided.

The change in eigenspectrum caused by the use of B-localization can be
better understood by noting the following (suggested by Dr. T. Tsuyuki;
private communication): as we saw in section \ref{sec:role-of-loc}, the
rank of $\mathbf{B}^\mathrm{ens}$, or eqivalently the number of non-zero
eigenvalues $\lambda^b_i$ of
$\mathbf{H}'\mathbf{B}^\mathrm{ens}\mathbf{H}'^T$, increases from $K-1$
to $KL-1$ by applying B-localization. Now, recalling that model-space
and observation-space B-localization are equivalent in this particular
case where $\mathbf{H}$ only picks up state variables at selected grid
points, the matrix
$\mathbf{H}'\left(\bm{\rho}\circ\mathbf{B}^\mathrm{ens}\right)\mathbf{H}'^T$
can be expressed as
$\bm{\rho}_\mathrm{obs}\circ\left(\mathbf{H}'\mathbf{B}^\mathrm{ens}\mathbf{H}'^T\right)$
by choosing an appropriate $p\times p$ correlation matrix
$\bm{\rho}_\mathrm{obs}$. Since all the diagonal entries of
$\bm{\rho}_\mathrm{obs}$ are one, it follows that
$\sum_{i=1}^{K-1}\lambda^{b,\mathrm{raw}}_i=\tr
\mathbf{H}'\mathbf{B}^\mathrm{ens}\mathbf{H}'^T =\tr
\bm{\rho}_\mathbf{obs}\circ\left(\mathbf{H}'\mathbf{B}^\mathrm{ens}\mathbf{H}'^T\right)
=\sum_{i=1}^{KL-1}\lambda^{b,\mathrm{loc}}_i$, where
$\lambda^{b,\mathrm{raw}}_i$ and $\lambda^{b,\mathrm{loc}}_i$ denote,
respectively, the eigenvalues of the prior error covariance matrix
measured in the normalized observation space before and after
application of B-localization. This means that the sum of all the
eigenvalues of the prior error covariance is invariant under application
of B-localization, while the number of its non-zero elements increases
$L$-fold, which implies that B-localization reduces the larger
eigenvalues while increasing the smaller eigenvalues including those
that are null, leading to flattening of the prior
eigenspectrum. Consequently, the posterior eigenspectrum is also
flattened by B-localization since the posterior eigenvalue $\lambda^a_i$
as a function of the prior eigenvalue $\lambda^b_i$ is monotonically
increasing (see Equation \ref{eq:eigval-A}), which explains how the
posterior eigenspectrum is deformed by B-localization from the gray
thick dotted line to the thin black solid line shown in Figure
\ref{fig:eig-HAH}.

\subsubsection{R-localization}
\label{sec:exp-role-of-R-loc}

R-localization circumvents this DFS underestimation issue in a different
manner. As we saw in section \ref{sec:role-of-loc}, with R-localization,
$\mbox{DFS}^\mathrm{ens}$ is inevitably smaller than the ensemble size
so DFS underestimation is unavoidable as long as
$\mbox{DFS}^\mathrm{opt}$ is greater than the ensemble size. Instead of
solving the full data assimilation problem using the full $\mathbf{B}$
matrix, localized EnKF algorithms like LETKF divide the domain into
smaller pieces exploiting the localized structure of the background
error covariance (\ie, the block diagonality of the $\mathbf{B}$ matrix)
and solves smaller data assimilation problems using the diagonal
submatrices of $\mathbf{B}$ \citep{Evensen2004,Ott2004}. The resultant
smaller data assimilation problems are solved independently for each
local domain. In our simple covariance model with the parameter $d=20$
(Figure \ref{fig:B-structure}b) , for example, the $\mathbf{B}$ matrix
assumes localized structure with almost zero correlations beyond $\sim
15$ grid points apart, so that, when performing analysis for a
particular grid point, limiting the domain to the neighboring grid
points within some radius (called cut-off distance hereafter) greater
than $\sim 15$ and then performing data assimilation, neglecting all
background field and observations outside this local domain, should
yield analysis close to what would be obtained with global analysis
using full background field and observations. This is exactly how data
assimilation is performed with LETKF. This way the number of locally
assimilated observations can be reduced \citep{Hunt2007}, resulting in
$\mbox{DFS}^\mathrm{opt}$ for the localized problem being smaller than
the ensemble size if the cut-off distance is chosen small enough, thus
allowing us to circumvent the DFS underestimation issue.

\begin{figure}[htbp]
 \centering
 \includegraphics[bb=0 0 590 400, width=0.45\textwidth]{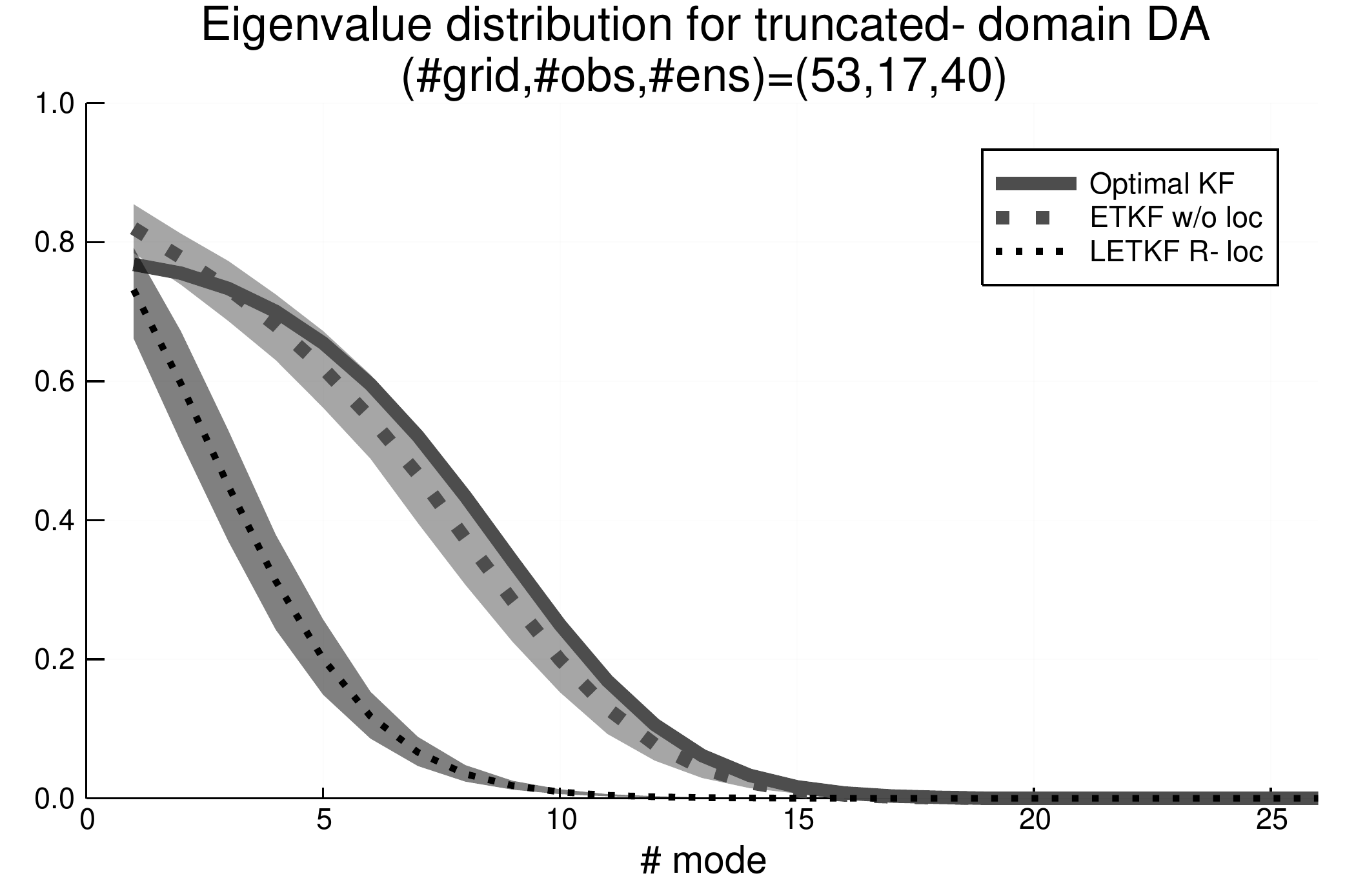} 
 \caption{
 As in Figure \ref{fig:eig-HAH}, but for a data assimilation problem for
 a smaller domain localized with a cut-off distance of 26. Shown are the
 eigenvalues of the matrix
 $\mathbf{H'}_\mathrm{loc}\mathbf{A}^\mathrm{opt}\mathbf{H'}^T_\mathrm{loc}$
 (gray thick solid line),
 $\mathbf{H'}_\mathrm{loc}\mathbf{A}^\mathrm{ens}\mathbf{H'}^T_\mathrm{loc}$
 (gray thick dotted line), and their counterpart for LETKF with
 R-localization (black thin dotted line). As in Figure
 \ref{fig:eig-HAH}, the shades represent the 5 to 95 percentile ranges.
 } \label{fig:eig-HAH-loc}
\end{figure}

In the case we are considering (with parameter $d=20$ and the ensemble
size $K=40$), the optimally-tuned cut-off distance for R-localization
was $d_\mathrm{cut-off}=14\sqrt{10/3}\approx 26$, in which case the
number of the locally assimilated observations is restricted to only
$17(\approx (2\times 26+1)/3)$. The gray thick solid line plotted in
Figure \ref{fig:eig-HAH-loc} shows the eigenvalues of the posterior
error covariance matrix projected onto the normalized observation space
$\mathbf{H'}_\mathrm{loc}\mathbf{A}_\mathrm{loc}^\mathrm{opt}\mathbf{H'}_\mathrm{loc}^T$
computed by the optimal canonical KF restricted to this localized domain
(here the matrix $\mathbf{H'}_\mathrm{loc}$ denotes a normalized
observation operator similar to $\mathbf{H'}$ but which selects only the
observations whose distances from the analyzed grid point are below the
cut-off distance). Here we arbitrarily show the result for the localized
problem centered around the first grid point but the results are
insensitive to the choice of the analyzed grid point because of the
translational symmetry of our covariance model. Note that the
eigenvalues are zero beyond the 18th mode since the rank of the matrix
$\mathbf{H'}_\mathrm{loc}\mathbf{A}_\mathrm{loc}^\mathrm{opt}\mathbf{H'}_\mathrm{loc}^T$
is only 17 (the number of locally assimilated observations). Their
ensemble equivalent computed with raw ensemble covariance but with the
domain localization (gray thick dotted line) is close to the optimal one
unlike in the case of full domain (compare with Figure
\ref{fig:eig-HAH}). When R-localization is applied, the posterior
eigenvalues become smaller than when only domain localization is applied
(black thin dotted line), meaning that R-localization actually leads to
smaller DFS than without. This is consistent with the finding of
\cite{Huang2019} who also provided a mathematical explanation.

\begin{figure}[htbp]
 \centering
 \includegraphics[bb=0 0 590 400, width=0.45\textwidth]{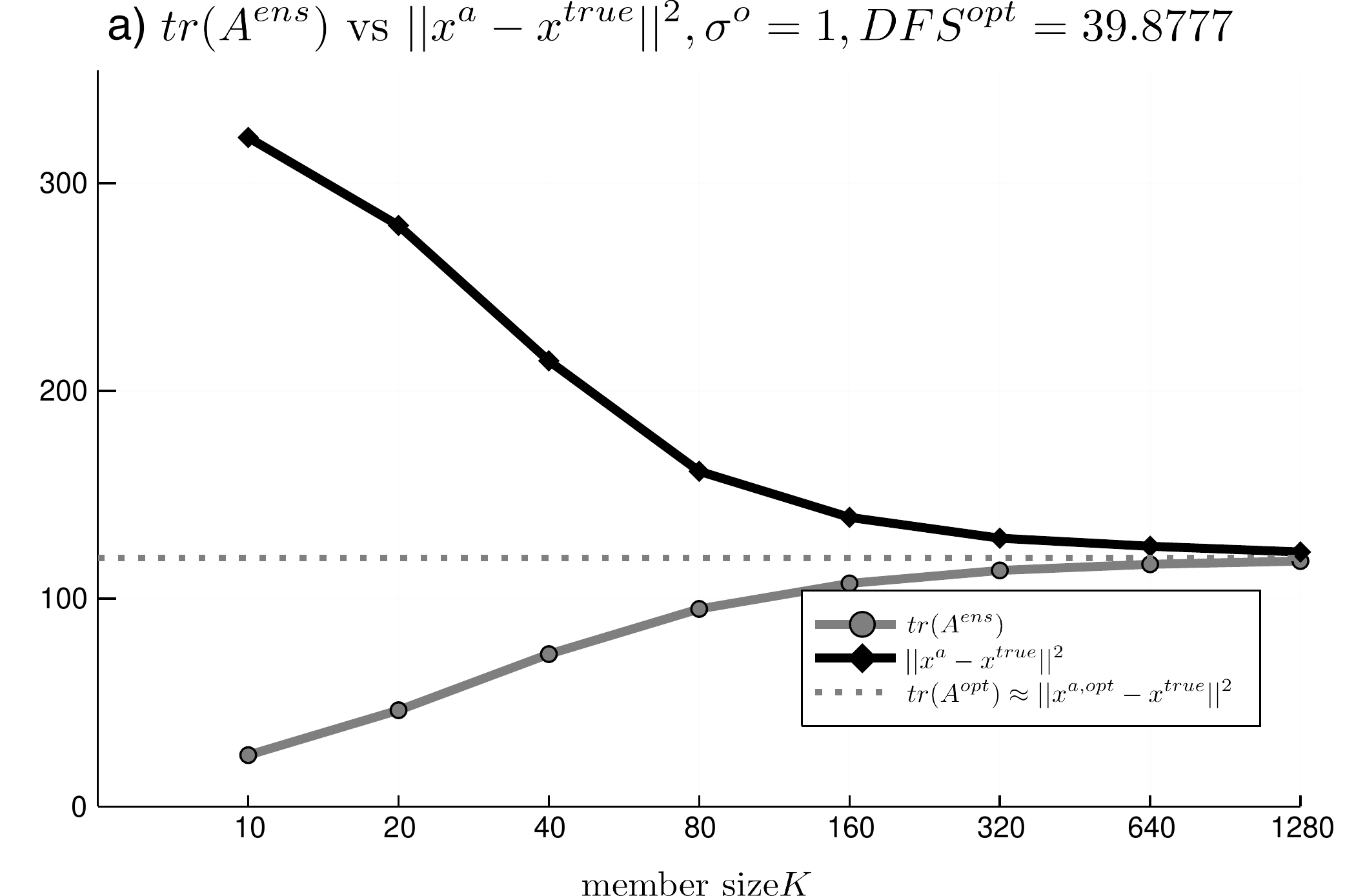}  
 \includegraphics[bb=0 0 590 400, width=0.45\textwidth]{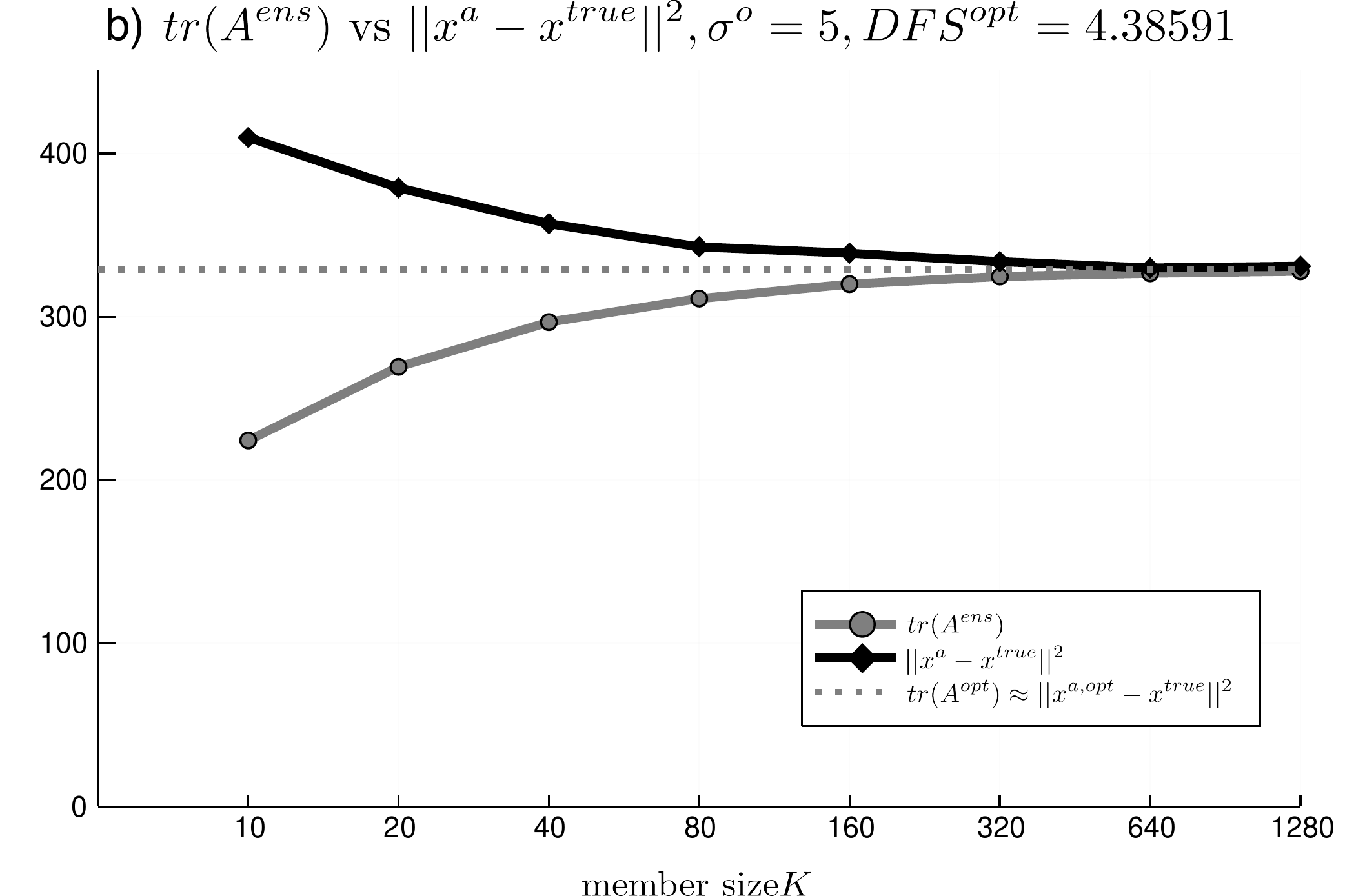}    
 \caption{
 Trace of analysis error covariance $\tr \mathbf{A}^\mathrm{ens}$ (gray
 filled circle) and the analysis MSE (black filled square) for ETKF
 analysis (or each local analysis of LETKF) plotted as a function of the
 ensemble size $K$, for (a) the ``high DFS'' and (b) ``low DFS''
 scenarios. As a reference, their counterpart obtained by an optimal
 canonical KF, is plotted in each panel as a horizontal dashed line.
 } \label{fig:trA-mse-vs-K}
\end{figure}

\begin{figure}[htbp]
 \centering
 \includegraphics[bb=0 0 590 400, width=0.45\textwidth]{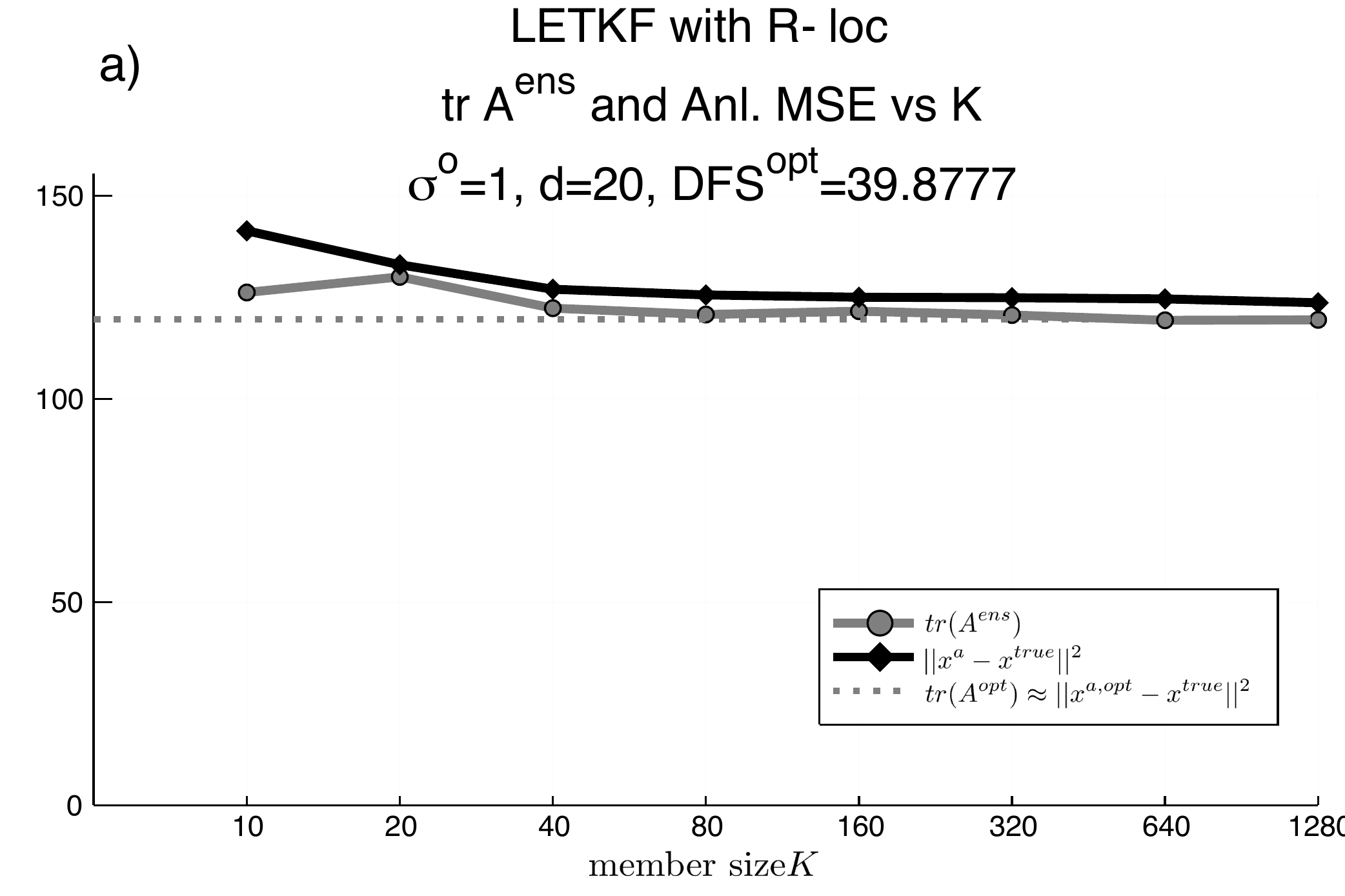}  
 \includegraphics[bb=0 0 590 400, width=0.45\textwidth]{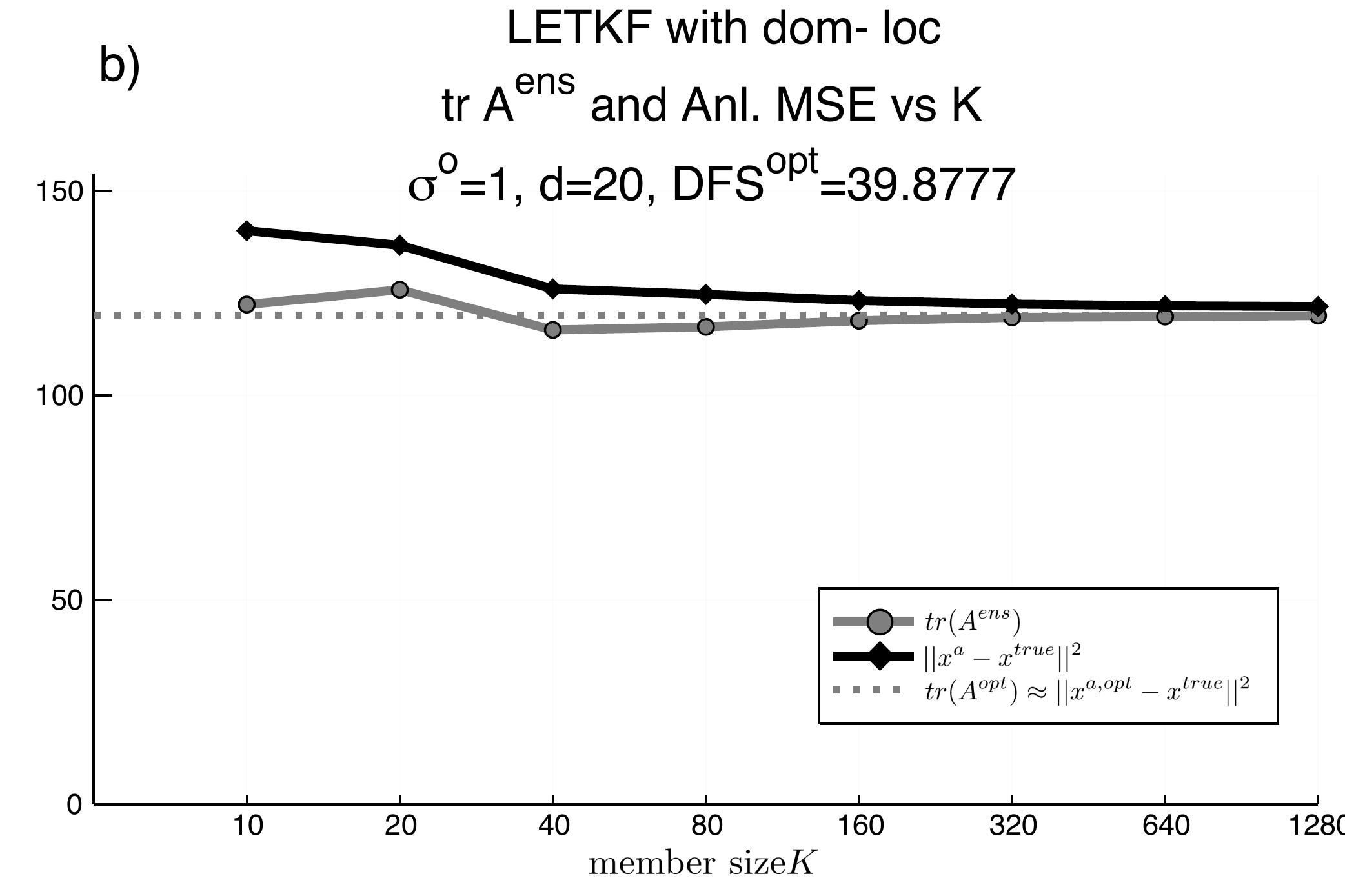}    
 \caption{
 As in Figure \ref{fig:trA-mse-vs-K}a, but for LETKF with (a)
 R-localization together with domain localization, and (b) domain
 localization only. For each ensemble size $K$, the localization
 parameter $d_\mathrm{cut-off}$ is manually tuned to yield the smallest
 analysis MSE with respect to the truth.
 } \label{fig:trA-mse-vs-K-loc}
\end{figure}

By working with the localized domain, the DFS underestimation is mostly
mitigated, and so is the overconfidence in the analysis spread. The
panels (a) and (b) in Figure \ref{fig:trA-mse-vs-K-loc} show,
respectively, a plot similar to Figure \ref{fig:trA-mse-vs-K}a but for
optimally-tuned LETKF with R-localization (together with its inherent
domain localization), and the same plot but with domain localization
only. Comparing these with Figure \ref{fig:trA-mse-vs-K}a, it is evident
that LETKF analyses suffer much less from overconfidence issue than the
raw ETKF without any localization (Figure \ref{fig:trA-mse-vs-K}a), and
that the analysis MSE close to that of the optimal KF is achieved with
much smaller ensemble sizes, proving the effectiveness of domain
localization. Interestingly, for this particular simple problem, the
benefit of R-localization appears to be mostly attributable to the use
of domain localization, as we can infer by comparing Figures
\ref{fig:trA-mse-vs-K-loc}a and \ref{fig:trA-mse-vs-K-loc}b which show
very similar performances. Note that we should not discredit the
advantage of R-localization over domain localization simply based on
these results. R-localization is known to have the advantage of ensuring
spatially smoother (and thus better balanced) analysis than simple
domain localization does. This advantage is particularly important in a
cycled context \citep{Greybush2011}, but is totally dismissed in our
problem setup.

In LETKF with R-localization or domain localization, the Kalman gain for
the global analysis in full domain (denoted $\mathbf{K}^\mathrm{LEKTF}$
hereafter), or the analysis error covariance implied by
$\mathbf{K}^\mathrm{LEKTF}$, are not explicitly available in algebraic
matrix form, but it is possible to numerically compute them row-wise if
we note that each row (say the $i$-th row) of the global Kalman gain
$\mathbf{K}^\mathrm{LEKTF}$ is computed in the local analysis centered
around the $i$-th grid point as the row of local Kalman gain that
corresponds to the center of this local analysis. The row from the local
Kalman gain is shorter than the corresponding row of the global Kalman
gain, so the components of the latter that corresponds to the
observations that are outside the truncated local domain have to be
padded with zero. Once $\mathbf{K}^\mathrm{LEKTF}$ is thus computed, a
counterpart of Figure \ref{fig:eig-HAH} can be produced, albeit with
several caveats to be kept in mind.

In the global analysis of LETKF, the optimality of the Kalman gain (in
the sense of Equation \ref{eq:Kgain-def} being strict), while being
valid for each local domain, does not globally hold exactly, which means
that Equation \ref{eq:K-AHtRinv} and thus Equation \ref{eq:HAH-def} are
not valid. Accordingly the equivalence of $\mathbf{H'AH'}^T$ and
$\mathbf{R}^{-1/2}\mathbf{HK}\mathbf{R}^{1/2}$ does not hold for the
global analysis, making it difficult to interpret the DFS (defined as
the trace of $\mathbf{HK}$) as the posterior error variance represented
by the assimilation scheme. Another difficulty that follows from
Equation \ref{eq:Kgain-def} being not exact is that
$\mathbf{R}^{-1/2}\mathbf{HK}^\mathrm{LETKF}\mathbf{R}^{1/2}$ is not
assured to be symmetric, resulting in their eigenvalues not necessarily real
values. An easy remedy to this is to focus on its symmetric component
\begin{align}
 \mathbf{H'K'}^{\mathrm{LETKF},S}:=
\frac{1}{2}\left\{
  \left(\mathbf{R}^{-1/2}\mathbf{HK}^\mathrm{LETKF}\mathbf{R}^{1/2}\right)+
  \left(\mathbf{R}^{-1/2}\mathbf{HK}^\mathrm{LETKF}\mathbf{R}^{1/2}\right)^T
\right\}
\end{align}
and investigate its eigenspectrum (suggested by Dr. T. Tsuyuki; private
communication).  This remedy may seem ad hoc but is justified by the following two properties:
Firstly, it conserves the trace and thus respects the identity 
\begin{align}
 \tr \mathbf{H'K'}^{\mathrm{LETKF},S}
=\tr \mathbf{R}^{-1/2}\mathbf{HK}^\mathrm{LETKF}\mathbf{R}^{1/2} 
= \tr \mathbf{HK}^\mathrm{LETKF} = \mbox{DFS}^\mathrm{ens}.
\end{align}
Secondly, by focusing on the symmetric component, the requirement that
eigenvalues $\lambda^a_i$ of $\mathbf{H'K'}^{\mathrm{LETKF},S}$ should
lie between 0 and 1 can be interpreted intuitively as requiring that the
analysis should be an interpolation between the first guess and the
observation. To see why the second point holds, recall that the
observation-space inner product (scaled by $\mathbf{R}^{-1}$) between
the innovation vector and the analysis increment can be expressed as
\begin{align}
&\left(\mathbf{y}^o-\mathbf{y}^b\right)^T\mathbf{R}^{-1}\left(\mathbf{y}^a-\mathbf{y}^b\right)
= \mathbf{d}^T\mathbf{R}^{-1}\left(\mathbf{y}^a-\mathbf{y}^b\right)
= \mathbf{d}^T\mathbf{R}^{-1}  \mathbf{HK}^\mathrm{LETKF}\mathbf{d} 
=\mathbf{d'}^T\mathbf{R}^{-1/2}\mathbf{HK}^\mathrm{LETKF}\mathbf{R}^{1/2}\mathbf{d'}  \nonumber \\
&=\frac{1}{2}\left\{
 \left(\mathbf{d'}^T\mathbf{R}^{-1/2}\mathbf{HK}^\mathrm{LETKF}\mathbf{R}^{1/2}\mathbf{d'}\right)
+\left(\mathbf{d'}^T\mathbf{R}^{-1/2}\mathbf{HK}^\mathrm{LETKF}\mathbf{R}^{1/2}\mathbf{d'}\right)^T
\right\}
=\mathbf{d'}^T \mathbf{H'K'}^{\mathrm{LETKF},S}\mathbf{d'}
\end{align}
where $\mathbf{y}^b:=\mathbf{Hx}^b$ is the first guess in the
observation space and $\mathbf{d'}:=\mathbf{R}^{-1/2}\mathbf{d}$ is the
normalized innovation vector. Since the eigenvalues $\lambda^a_i$ are
the factors by which the vector $\mathbf{d'}$ magnifies or shrinks in
the corresponding eigen-directions when multiplied from left by
$\mathbf{H'K'}^{\mathrm{LETKF},S}$, if $0<\lambda^a_i<1$ holds for all
$i$, that implies that, taking $\mathbf{y}^b$ as the origin in the
normalized observation space, in any direction, $\mathbf{y}^a$ never
goes beyond $\mathbf{y}^o$ nor does it lie in the opposite side of
$\mathbf{y}^o$ with respect to the origin $\mathbf{y}^b$, or stated
differently, $\mathbf{y}^a$ is an interpolation between $\mathbf{y}^b$
and $\mathbf{y}^o$.

\begin{figure}[htbp]
 \centering
 \includegraphics[bb=0 0 590 400, width=0.45\textwidth]{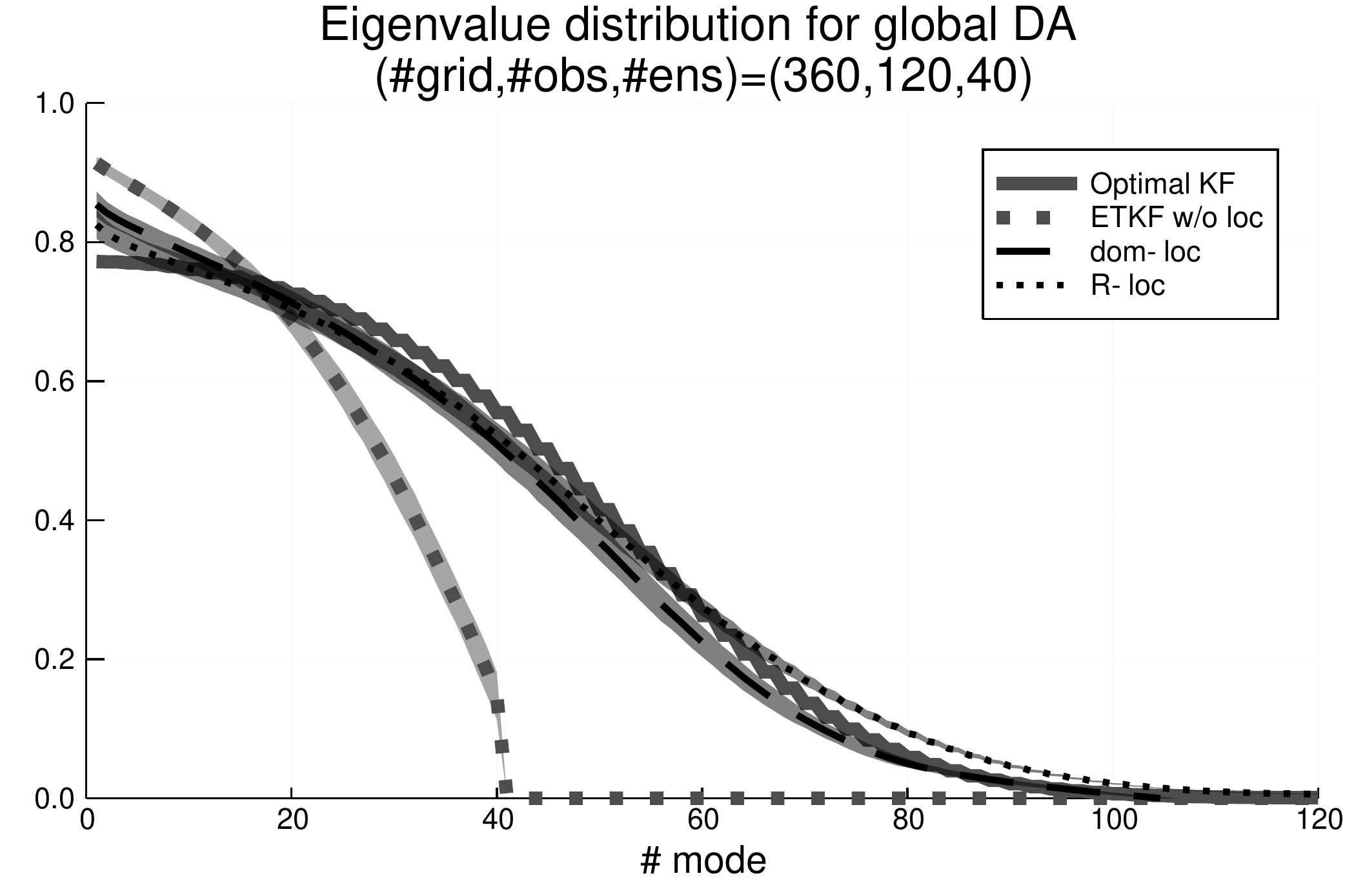} 
 \caption{
 As in Figure \ref{fig:eig-HAH}, but for the global analysis of LETKF
 with the optimally tuned R-localization (black thin dotted line) and
 LETKF with domain localization only (black thin dashed line). The
 latter uses the same cut-off distance $d_\mathrm{cut-off}=26$ as the
 former. For comparison, the posterior eigenspectra of the optimal KF
 (gray thick solid line) and the raw ETKF without localization (gray
 thick dotted line) are plotted here again. As in Figure
 \ref{fig:eig-HAH}, the shades represent the 5 to 95 percentile ranges.
 } \label{fig:eig-HAH-Rloc}
\end{figure}

The eigenspectrum of $\mathbf{H'K'}^{\mathrm{LETKF},S}$, computed with
the optimally-tuned R-localization or with the domain localization only,
are plotted in Figure \ref{fig:eig-HAH-Rloc}. Similarly to
B-localization (see Figure \ref{fig:eig-HAH}), R-localization and domain
localization successfully restore the true eigenspectrum (gray thick
solid line), attesting the effectiveness of R-localization or domain
localization in the context of global analysis. Interestingly again, the
two curves (for R-localization and domain localization) exhibit very
similar pattern, suggesting that much of the benefit of R-localization
can be achieved by domain localization alone. 
Since the underestimation of $\tr \mathbf{HK}$ (the area below the
curves) is alleviated, we can expect that R-localization or domain
localization should alleviate the underestimation of the analysis
increment ($\mathbf{HKd}$ or $\mathbf{Kd}$). We confirmed that this is
indeed the case in our experiment: the ratio of the $l_2$-norm of
analysis increment computed by ETKF (averaged over 1,000 trials) divided
by the same quantity computed by the optimal canonocal KF increased from
0.7 to 0.9.  
We remark that, since the identity between
$\mathbf{H'K'}^{\mathrm{LETKF},S}$ and $\mathbf{H'AH'}^T$ does not
necessarily hold for global analysis of LETKF, the mitigation of DFS
(the sum of all $\lambda^a_i$) cannot be immediately interpreted as
mitigation of posterior underdispersion (although from Figure
\ref{fig:trA-mse-vs-K-loc} it does seem quite likely to be mitigated).

\section{Discussions: interpretation of results from the literature in light of DFS} 
\label{sec:discussion}

In the literature of atmospheric data assimilation studies, especially
those on operational implementation of LETKF, several interesting
findings have been reported and some of them appear to be
counter-intuitive. This section is devoted to discussing how DFS-based
arguments developed in the preceding sections could help interpret such
seemingly puzzling findings. The discussions given here are admittedly
highly speculative, but the authors wish nonetheless to present them in
hopes of stimulating further discussions by the community.

\subsection{Using less is better}
\label{sec:discussion-less-is-good}

Among the many findings (or caveats) on operational or real-world
implementations of LETKF, perhaps the most intriguing is the fact that
assimilating less observations can lead to more accurate analysis or
forecast. For example, \cite{Hamrud2015} developed quasi-operational
implementation of both LETKF and serial ensemble square-root filter
\citep[EnSRF][]{WhitakerHamill2002} coupled with the European Centre for
Medium-range Weather Forecasts (ECMWF)'s operational forecast model
(Integrated Forecast System; IFS) and found that, for LETKF, a very
large forecast improvement is achieved by severely limiting the number
of assimilated observations in the local analysis step for each analyzed
grid point. They reported that, reducing the average number of locally
assimilated observations $\sim 30$-fold from the original $\sim 40,000$
to only $\sim 1,200$ yields the best results and that the forecast
performance is not very sensitive to the exact choice of the strength of
number reduction. Curiously, they found this method (which they call
``implicit covariance localization'') to be useful only for LETKF (with
R-localization) and not for serial EnSRF (with observation-space
B-localization).

The LETKF implementation for convective-scale data assimilation
developed by \cite{Schraff2016} takes a similar approach where, for each
local analysis, the horizontal localization length-scale is adjusted so
that the number of locally assimilated observations become constant
(roughly double the ensemble size) at every analyzed grid point.
Guo-Yuan Lien et al. (2017; private communication) applied a similar
method to assimilation of phased array weather radar (PAWR) data by
their LETKF \citep{Lien2017} and confirmed that limiting the number of
locally assimilated observations to a few times the ensemble size leads
to significantly better forecast than assimilating all data or applying
a traditional thinning method before assimilating them.

The fact that using less observations (\ie, discarding many of the
available observations) leads to better forecast performance is,
naively, not easily justifiable, but the theory of DFS applied to EnKF
(see section \ref{sec:dfs-ens}) allows us a clear interpretation:
$\mbox{DFS}^\mathrm{ens}$ is locally at most $K – 1$ (where $K$ is the
ensemble size), so that locally assimilating much more than $K$
observations results in overconfident analysis spread (requiring
unreasonably large inflation) and smaller-than-optimal analysis
increment. The above interpretation is not necessarily new, and we note
that \cite{Schraff2016} presents a similar heuristic argument as a
rationale for their approach.

One may argue that the benefit of assimilating less observations should
be attributable to the observations being sparser and thus less affected
by the error correlation between different observations. However, such
an interpretation seems not to apply here, because, in all of the three
studies mentioned above, the observations that are spatially closest to
the analyzed grid point are selected, so that the issue of correlated
observation errors, if any, was not addressed by limiting the number of
locally assimilated observations.

\subsection{Relationship between the optimal localization scale and the ensemble size}

The argument above suggests a convenient guidance on how to tune the
localization scale in the context of LETKF with R-localization: \\ 
the localization scale should be as large as possible to keep the
localized problem close to the original global problem, but {\it subject
to the condition that it is small enough so that the number of locally
assimilated observations does not exceed several times the ensemble
size} (to ensure that locally $\mbox{DFS}^\mathrm{opt} < K$ holds).

This rule of thumb is again not new, and the insightful review paper by
\cite{Tsyrulnikov2010} developed a similar heuristic argument based in
part on \cite{Lorenc2003} to reach at a similar conclusion. Quoting from
section 4.3 of \cite{Tsyrulnikov2010}, he reasoned that the optimal
localization scale occurs when local analysis domain is small enough so
that ``ensemble size (is) commensurable with the number of observed
degrees of freedom within an effective box.'' In his discussion, the
``observed degrees of freedom'' has not been given a precise definition;
we assert that DFS gives it a more precise definition.
\cite{Tsyrulnikov2010} further conducted meta-analysis of the published
literature and found that the optimal localization scales that were
experimentally determined in earlier studies do match this rule of
thumb.

Covariance localization has traditionally been regarded as a means to
alleviate the rank deficiency issue of the ensemble background error
covariance matrix $\mathbf{B}^\mathrm{ens}$
\citep[e.g.,][]{HoutekamerZhang2016} or a means to damp the sampling
noises of $\mathbf{B}^\mathrm{ens}$\citep[e.g.,][]{Menetrier2015}. From
this perspective, it appears that determination of the optimal
localization length scale is a problem intrinsic to the ensemble size
$K$ and the structure of the true $\mathbf{B}$, independent of
$\mathbf{H}$ or $\mathbf{R}$ (\ie, how the observations are distributed
or how they are accurate). Interestingly, however, contrary to this
traditional view, there have been ample evidence from the literature
that suggests that the distribution and/or accuracy of observations are
the key in determining the optimal localization scales in the context of
LETKF with R-localization and domain localization. The theory based on
DFS presented in this paper may help to reconcile this apparent
contradiction.

\subsection{Covariance inflation}
\label{sec:discussion-inflation}
Along with localization, covariance inflation is an indispensable
component of practical EnKF implementation that is often performed in an
ad hoc manner despite its importance. The concept of DFS is useful in
explaining/justifying some of the inflation methods that were found
effective in previous literature.

\cite{Wang2007} experimentally observed that the regular ETKF \citep[as
formulated in][]{Bishop2001,Wang2004} systematically underestimates the
posterior error variance if $K \ll r$ where $K$ is the ensemble size and
$r$ is the rank of the true background error covariance
$\mathbf{H'}\mathbf{B}^\mathrm{true}\mathbf{H'}^T$ projected onto the
normalized observation space. Motivated by this observation, and guided
by a series of insightful educated guesses (see their Appendix A),
\cite{Wang2007} introduced an ``improved ETKF'' formulation which
inflates the eigenvalues $\lambda^b_i$ of the sample prior error
covariance projected onto the normalized observation space before
computing the ensemble transform matrix. The inflation factor they
derived has a rather complex expression but its key property is that it
approaches $r/K$ (the rank of
$\mathbf{H'}\mathbf{B}^\mathrm{true}\mathbf{H'}^T$ divided by the
ensemble size) under some assumptions. The chain of logic behind their
derivation is somewhat complicated, but the DFS argument given in
section \ref{sec:implications-dfs} of this manuscript allows us an
intuitive (albeit heuristic) interpretation:
when $K \ll r$, DFS (and thus the posterior error variance measured in
the normalized observation space) should be underestimated
($\mbox{DFS}^\mathrm{ens} \ll \mbox{DFS}^\mathrm{opt}$) since
$\mbox{DFS}^\mathrm{ens} < K-1 \ll r$ and $r/\mbox{DFS}^\mathrm{opt}
\sim O(1)$ if observations are accurate enough. We can recover the
correct posterior variance by inflating $\mbox{DFS}^\mathrm{ens}$ by a
factor $\mbox{DFS}^\mathrm{opt}/\mbox{DFS}^\mathrm{ens}$, and a simple
way to do this is to inflate all the prior eigenvalues $\lambda^b_i$ by
the same factor $\mbox{DFS}^\mathrm{opt}/\mbox{DFS}^\mathrm{ens}$. This
factor is difficult to estimate since $\mbox{DFS}^\mathrm{opt}$ is
usually unknown (in fact \cite{Wang2007} derived quite an elaborate
expression to estimate this factor), but provided that $K \ll r$, it
should be reasonable to assume that it is roughly proportional to
$r/K$.

As we mentioned in section \ref{sec:role-of-loc}, \cite{BWL17} proposed
the Gain-form ETKF (GETKF) that enables model-space B-localization in
the ETKF framework. They observed that GETKF tend (though not always) to
underestimate the posterior error variance in comparison to METKF (the
ETKF that updates all the modulated ensemble members) and proposed to apply
the ``inherent GETKF inflation'' which inflates each of the GETKF's
posterior perturbations by a constant factor $a$ that restores the
model-space posterior variance of METKF. With experiments using a
one-dimesional toy system repeated with various localization length
scales, they found that:
\begin{itemize}
 \item (a) the inherent inflation factor $a$ increases monotonically with
       the localization scale, and
 \item (b) interestingly, analysis becomes most accurate when the
       localization length scale is such that it neutralizes the
       inherent inflation factor (\ie, $a\approx 1$ holds).
\end{itemize}
The point (a) above is easier to understand if we note, from the DFS
theory, that the larger the localization scale is, the more observations
are assimilated, leading to severer DFS (and thus posterior variance)
underestimation, requiring stronger covariance inflation. \cite{BWL17}
acknowledges the potential significance of point (b) suggesting that it
could be exploited to adaptively optimize localization scale if it is a
generally applicable property. \cite{BWL17} were deliberate in stating
that the validity of this hypothesis is left to future assessment, but
the implication discussed in section \ref{sec:implications-dfs} together
with the similar evidence from literature summarized in
\cite{Tsyrulnikov2010}, make it all the more likely. 

Finally, among the many inflation methods, a family of ``relaxation to
prior'' approaches have been found to be particularly effective. This
family of inflation methods modify (inflate) the posterior ensemble
after the analysis update has been made by relaxing the posterior
perturbations themselves to the prior perturbations \cite[Relaxation to
Prior Perturbations; RTPP][]{Zhang2004} or by relaxing the posterior
spread to the prior spread \citep[Relaxation to Prior Spread;
RTPS]{WhitakerHamill2012}. While there can be many reasons why these
relaxation approaches are effective, one important characteristic of
them appears to be their ability to apply stronger inflation when and
where the assimilated observations are distributed more densely, as
pointed out by \cite{WhitakerHamill2012}. The DFS underestimation (or
analysis overconfidence) that occurs when assimilating much more
observations than the ensemble size, lends itself to justify the success
of the relaxation approaches.

\section{Summary and concluding remarks} 
\label{sec:conclusions}

Aiming at understanding how EnKF effectively extracts information from
observations, in this paper we adapted the theory of DFS to EnKF
algorithms with particular focus on the ETKF framework. Simple
mathematical arguments based on elementary linear algebra revealed that,
with EnKF algorithms, DFS is bounded from above by the ensemble size,
which means that DFS is always underestimated when assimilating much
more observations than the available ensemble size. This problem has
long been recognized by the community and is referred to as the ``rank
deficiency issue'' but it appears to have been rather vaguely
defined. DFS argument allows us to describe this issue in a more
quantitative manner.

The fact that DFS is underestimated when assimilating much more
observations than the ensemble size has several important implications
on the effectiveness of practical EnKF implementations, notably:
\begin{itemize}
 \item Strong covariance inflation is necessary when assimilating much
       more observations than the ensemble size, which follows from the
       fact that DFS coincides with the analysis (posterior) variance
       measured in the normalized observation space, so that, if DFS is
       underestimated, the analysis spread becomes underdispersive (or
       equivalently, analysis becomes overconfident).
 \item DFS underestimation (or analysis overconfidence) can be avoided
       by imposing stronger localization when/where observations are
       denser. This comes at the expense of discarding some of the
       information from observations, but the merit of alleviating the
       overconfidence in analysis can outweigh such disadvantage.
\end{itemize}
These findings are not new, and similar arguments have been repeatedly
made in the literature
\citep[e.g.,][]{Lorenc2003,Tsyrulnikov2010,FurrerBengtsson2007}, but the
authors believe that the DFS-based argument presented here helps us to
understand the issue more clearly and more quantitatively.

The concept of DFS also allows us to understand how different
localization schemes help to circumvent the DFS underestimation or
analysis overconfidence. The implications from the DFS theory in this
context have been explored and showcased using idealized experiments
with a one-dimensional toy problem.

The examination on the role of localization based on the DFS concept
highlights the advantage of B-localization over R-localization (\ie,
being less susceptible to the problem of DFS underestimation or analysis
overconfidence). In the previous studies, the benefit of using
model-space B-localization, through the modulated ensemble approach in
particular, have been emphasized in connection to its ability to
correctly account for non-local observations like satellite radiances
for which the position of the observation is ill-defined and thus
R-localization cannot be clearly formulated \citep{BWL17,LWB18}. The
discussion given in this paper suggests the merit of B-localization
beyond its intended advantage of correctly accounting for non-locality
of observations.

Finally, several speculative arguments have been presented based on DFS
theory about how some of the interesting results on (L)ETKF reported in
the previous literature can be explained or justified in light of DFS
concept. Intriguing questions such as why using less observations can be
better, which localization length scales tend to be optimal, and why
covariance inflation schemes based on ``relaxation to prior'' approaches
are successful, all become easier to interpret by noting when DFS
underestimation issue occurs. We remark that the theoretical argument
and the discussions on the results from the idealized experiments
presented in this paper only bring out the issues that (L)ETKF
algorithms are subject to even when strong simplifying assumptions are
satisfied, such as: the model and observation operator are both linear,
the background and observation errors obey Gaussian distributions, the
observation errors are uncorrelated, and $\mathbf{B}$ and $\mathbf{R}$
are perfectly known. Practical applications like real-world NWP
bear many other complicating factors, so we need to be careful not to
extrapolate too much from the simple theoretical arguments presented in
this paper. The authors believe nevertheless that the findings shown
here provide some useful insights.

Most DA methodologies studied in the atmospheric science or geophysics
literature have assumed, implicitly or explicitly, that the system
dimension is by far greater than the number of observations or the
ensemble size (\#ens $\sim$ \#obs $\ll$ \#grid) often emphasizing the
underdetermined nature of DA problems.  With the advent of
meteorological Big Data, however, this assumption will no longer be
justifiable. The situation that we should consider now is cases where
there are about as many observations as the system dimension which by
far exceed the affordable member size (\ie, \#ens $\ll$ \#obs $\sim$
\#grid). The authors hope that the DFS concept will prove useful in
developing DA methods suitable for this new emerging (or incoming)
situation.

\section*{ACKNOWLEDGEMENTS}
This paper grew out from the first author's presentations at the 6th
International Symposium on Data Assimilation (ISDA), the 7th and 8th
EnKF Data Assimilation Workshops and the RIKEN International Symposium
on Data Assimilation (RISDA2017). The authors are grateful to the
organizers and thank feedback from many of the participants, notably (in
no particular order), Prof. Craig Bishop (University of Melbourne),
Prof. Eugenia Kalnay (University of Maryland), Dr. Takemasa Miyoshi
(RIKEN), Dr. Guo-Yuan Lien (CWB), Dr. Jeff Whitaker (NOAA/ESRL),
Prof. Shunji Kotsuki (Chiba University), Prof. Kosuke Ito (University of
the Ryukyus) and Dr. Le Duc (JAMSTEC). The authors also wish to thank
Dr. Tadashi Tsuyuki (MRI) for reviewing the pre-submission version of
the manuscript and for providing numerous insightful suggestions.
This work is supported in part by JSPS Grant-in-Aid for Scientific
Research (KAKENHI) JP17H00852 and JP17H07352, the MEXT FLAGSHIP2020
Project within the priority study 4 (advancement of meteorological and
global environmental predictions utilizing observational Big Data), and
the MEXT “Program for Promoting Researches on the Supercomputer
Fugaku” (Large Ensemble Atmospheric and Environmental Prediction for
Disaster Prevention and Mitigation).

\section*{CONFLICT OF INTEREST}

Authors declare no conflict of interest.



\begin{thebibliography}{46}
\expandafter\ifx\csname natexlab\endcsname\relax\def\natexlab#1{#1}\fi
\expandafter\ifx\csname url\endcsname\relax
  \def\url#1{\texttt{#1}}\fi
\expandafter\ifx\csname urlprefix\endcsname\relax\def\urlprefix{URL: }\fi

\bibitem[{Anderson(2001)}]{Anderson2001}
Anderson, J. (2001) {An ensemble adjustment Kalman filter for data
  assimilation}.
\newblock \textit{Monthly Weather Review}, \textbf{129}, 2884–2903.

\bibitem[{Anderson and Anderson(1999)}]{AndersonAnderson1999}
Anderson, J. and Anderson, S. (1999) {A Monte Carlo implementation of the
  nonlinear filtering problem to produce ensemble assimilations and forecasts}.
\newblock \textit{Monthly Weather Review}, \textbf{127}, 2741–2758.

\bibitem[{Bishop et~al.(2001)Bishop, Etherton and Majumdar}]{Bishop2001}
Bishop, C.~H., Etherton, B.~J. and Majumdar, S.~J. (2001) {Adaptive sampling
  with the ensemble transform Kalman filter. Part I: Theoretical aspects}.
\newblock \textit{Monthly Weather Review}, \textbf{129}, 420–436.

\bibitem[{Bishop and Hodyss(2009{\natexlab{a}})}]{BishopHodyss2009a}
Bishop, C.~H. and Hodyss, D. (2009{\natexlab{a}}) Ensemble covariances
  adaptively localized with {ECO-RAP}. {Part 1}: Tests on simple error models.
\newblock \textit{Tellus}, \textbf{61A}, 84---96.

\bibitem[{Bishop and Hodyss(2009{\natexlab{b}})}]{BishopHodyss2009b}
--- (2009{\natexlab{b}}) Ensemble covariances adaptively localized with
  {ECO-RAP}. {Part 2}: A strategy for the atmosphere.
\newblock \textit{Tellus}, \textbf{61A}, 97--111.

\bibitem[{Bishop et~al.(2017)Bishop, Whitaker and Lei}]{BWL17}
Bishop, C.~H., Whitaker, J.~S. and Lei, L. (2017) Gain form of the {Ensemble
  Transform Kalman Filter} and its relevance to satellite data assimilation
  with model space ensemble covariance localization.
\newblock \textit{Monthly Weather Review}, \textbf{145}, 4574--4592.

\bibitem[{Bocquet(2016)}]{Bocquet2016}
Bocquet, M. (2016) Localization and the iterative ensemble kalman smoother.
\newblock \textit{Quartely Journal of the Royal Meteorological Society},
  \textbf{142}, 1075--1089.

\bibitem[{Bocquet et~al.(2011)Bocquet, Wu and Chevallier}]{Bocquet2011}
Bocquet, M., Wu, L. and Chevallier, F. (2011) {Bayesian design of control space
  for optimal assimilation of observations. I: Consistent multiscale
  formalism}.
\newblock \textit{Quartely Journal of the Royal Meteorological Society},
  \textbf{137}, 1340–1356.

\bibitem[{Cardinali(2004)}]{Cardinali2004}
Cardinali, C. (2004) Monitoring the observation impact on the short-range
  forecast.
\newblock \textit{Quartely Journal of the Royal Meteorological Society},
  \textbf{135}, 239--250.

\bibitem[{Cardinali(2013)}]{Cardinali2013}
--- (2013) Observation influence diagnostic of a data assimilation system.
\newblock In \textit{Data Assimilation for Atmospheric, Oceanic and Hydrology
  Applications} (ed. X.~P. Park~SK), vol.~2, chap.~4, 89–110. Berlin and
  Heidelberg, Germany: Springer‐Verlag.

\bibitem[{Chapnik et~al.(2006)Chapnik, Desroziers, Rabier and
  Talagrand}]{Chapnik2006}
Chapnik, B., Desroziers, G., Rabier, F. and Talagrand, O. (2006) Diagnosis and
  tuning of observational error in a quasi-operational data assimilation
  setting.
\newblock \textit{Quartely Journal of the Royal Meteorological Society},
  \textbf{132}, 543–565.

\bibitem[{Evensen(2004)}]{Evensen2004}
Evensen, G. (2004) Sampling strategies and square root analysis schemes for the
  {EnKF}.
\newblock \textit{Ocean Dynamics}, \textbf{54}, 539--560.

\bibitem[{Fisher(2003)}]{Fisher2003}
Fisher, M. (2003) Estimation of entropy reduction and degrees of freedom for
  signal for large variational analysis systems.
\newblock \textit{ECMWF Technical Memoranda}, \textbf{397}, 18pp.

\bibitem[{Furrer and Bengtsson(2007)}]{FurrerBengtsson2007}
Furrer, R. and Bengtsson, T. (2007) {Estimation of high-dimensional prior and
  posterior covariance matrices in Kalman filter variants}.
\newblock \textit{Journal of Multivariate Analysis}, \textbf{98}, 227--255.

\bibitem[{Gaspari and Cohn(1999)}]{GaspariCohn1999}
Gaspari, G. and Cohn, E. (1999) Construction of correlation functions in two
  and three dimensions.
\newblock \textit{Quartely Journal of the Royal Meteorological Society},
  \textbf{125}, 723--757.

\bibitem[{Greybush et~al.(2011)Greybush, Kalnay, Miyoshi, Ide and
  Hunt}]{Greybush2011}
Greybush, S.~J., Kalnay, E., Miyoshi, T., Ide, K. and Hunt, B.~R. (2011)
  Balance and ensemble kalman filter localization techniques.
\newblock \textit{Monthly Weather Review}, \textbf{139}, 511–522.

\bibitem[{Hamrud et~al.(2015)Hamrud, Bonavita and Isaksen}]{Hamrud2015}
Hamrud, M., Bonavita, M. and Isaksen, L. (2015) {EnKF} and hybrid gain ensemble
  data assimilation. {Part} i: {EnKF} implementation.
\newblock \textit{Monthly Weather Review}, \textbf{143}, 4847--4864.

\bibitem[{Houtekamer and Zhang(2016)}]{HoutekamerZhang2016}
Houtekamer, P.~L. and Zhang, F. (2016) Review of the {Ensemble Kalman Filter}
  for atmospheric data assimilation.
\newblock \textit{Monthly Weather Review}, \textbf{144}, 4489--4532.

\bibitem[{Huang et~al.(2019)Huang, Wang and Bishop}]{Huang2019}
Huang, B., Wang, X. and Bishop, C.~H. (2019) {The High-rank Ensemble Transform
  Kalman Filter}.
\newblock \textit{Monthly Weather Review}, \textbf{147}, 3025–3043.

\bibitem[{Hunt et~al.(2007)Hunt, Kostelich and Szunyogh}]{Hunt2007}
Hunt, B., Kostelich, E. and Szunyogh, I. (2007) {Efficient data assimilation
  for spatiotemporal chaos: A local ensemble transform Kalman filter}.
\newblock \textit{Physica D}, \textbf{230}, 112–126.

\bibitem[{JMA(2013)}]{JMA-outline13}
JMA (2013) Outline of the operational numerical weather prediction at the
  {Japan Meteorological Agency (March 2013)}, {Appendix to WMO Technical
  Progress Report on the Global Data-processing and Forecasting System (GDPFS)
  and Numerical Weather Prediction (NWP) Research}.
\newblock 188pp.
\newblock
  \urlprefix\url{http://www.jma.go.jp/jma/jma-eng/jma-center/nwp/outline2013-nwp/index.htm}.

\bibitem[{JMA(2019)}]{JMA-outline19}
--- (2019) Outline of the operational numerical weather prediction at the
  {Japan Meteorological Agency (March 2019)}, {Appendix to WMO Technical
  Progress Report on the Global Data-processing and Forecasting System (GDPFS)
  and Numerical Weather Prediction (NWP) Research}.
\newblock 229pp.
\newblock
  \urlprefix\url{http://www.jma.go.jp/jma/jma-eng/jma-center/nwp/outline2019-nwp/index.htm}.

\bibitem[{Johnson et~al.(2005)Johnson, Hoskins and Nichols}]{Johnson2005}
Johnson, C., Hoskins, B.~J. and Nichols, N.~K. (2005) {A singular vector
  perspective of 4D-Var: Filtering and interpolation}.
\newblock \textit{Quartely Journal of the Royal Meteorological Society},
  \textbf{131}, 1--19.

\bibitem[{Lei et~al.(2018)Lei, Whitaker and Bishop}]{LWB18}
Lei, L., Whitaker, J.~S. and Bishop, C. (2018) Improving assimilation of
  radiance observations by implementing model space localization in an ensemble
  {Kalman} filter.
\newblock \textit{Journal of Advances in Modeling Earth Systems},
  \textbf{10:12}, 3221--3232.

\bibitem[{Lien et~al.(2017)Lien, Miyoshi, Nishizawa, Yoshida, Yashiro, Adachi,
  Yamaura and Tomita}]{Lien2017}
Lien, G.-Y., Miyoshi, T., Nishizawa, S., Yoshida, R., Yashiro, H., Adachi,
  S.~A., Yamaura, T. and Tomita, H. (2017) {The near-real-time SCALE-LETKF
  system: A case of the September 2015 Kanto-Tohoku heavy rainfall}.
\newblock \textit{Scientific Online Letters on the Atmosphere}, \textbf{13},
  1–6.

\bibitem[{Liu et~al.(2009)Liu, Kalnay, Miyoshi and Cardinali}]{Liu2009}
Liu, J., Kalnay, E., Miyoshi, T. and Cardinali, C. (2009) Analysis sensitivity
  calculation in an ensemble kalman filter.
\newblock \textit{Quartely Journal of the Royal Meteorological Society},
  \textbf{135}, 1842–1851.

\bibitem[{Lorenc(2003)}]{Lorenc2003}
Lorenc, A.~C. (2003) The potential of the ensemble {Kalman} filter for {NWP}
  --- a comparison with {4D-Var}.
\newblock \textit{Quartely Journal of the Royal Meteorological Society},
  \textbf{129}, 3183--3203.

\bibitem[{Lupu et~al.(2011)Lupu, Gauthier and Laroche}]{Lupu2011}
Lupu, C., Gauthier, P. and Laroche, S. (2011) Evaluation of the impact of
  observations on analyses in 3d- and 4d-var based on information content.
\newblock \textit{Monthly Weather Review}, \textbf{139}, 726--737.

\bibitem[{M\'{e}n\'{e}trier et~al.(2015)M\'{e}n\'{e}trier, Montmerle, Michel
  and Berre}]{Menetrier2015}
M\'{e}n\'{e}trier, B., Montmerle, T., Michel, Y. and Berre, L. (2015) {Linear
  filtering of sample covariances for ensemble-based data assimilation. Part I:
  Optimality criteria and applications to variance filtering and covariance
  localization}.
\newblock \textit{Monthly Weather Review}, \textbf{143}, 1622--1643.

\bibitem[{Mitchell and Houtekamer(2000)}]{MitchellHoutekamer2000}
Mitchell, H.~L. and Houtekamer, P. (2000) {An adaptive ensemble Kalman filter}.
\newblock \textit{Monthly Weather Review}, \textbf{128}, 416–433.

\bibitem[{Miyoshi et~al.(2016)Miyoshi, Kunii, Ruiz, Lien, Satoh, Ushio, Bessho,
  Seko, Tomita and Ishikawa}]{Miyoshi2016}
Miyoshi, T., Kunii, M., Ruiz, J., Lien, G.-Y., Satoh, S., Ushio, T., Bessho,
  K., Seko, H., Tomita, H. and Ishikawa, Y. (2016) “big data assimilation”
  revolutionizing severe weather prediction.
\newblock \textit{Bulletin of the American Meteorological Society},
  \textbf{97}, 1347--1354.

\bibitem[{Ott and Coauthors(2004)}]{Ott2004}
Ott, E. and Coauthors (2004) A local ensemble {Kalman} filter for atmospheric
  data assimilation.
\newblock \textit{Tellus}, \textbf{56A}, 415--428.

\bibitem[{Pham et~al.(1998)Pham, Verron and Roubaud}]{Pham1998}
Pham, D.~T., Verron, J. and Roubaud, M.~C. (1998) A singular evolutive extended
  {Kalman} filter for data assimilation in oceanography.
\newblock \textit{Journal of Marine systems}, \textbf{16}, 323–340.

\bibitem[{Purser and Huang(1993)}]{PurserHuang1993}
Purser, R.~J. and Huang, H.~L. (1993) Estimating effective data density in a
  satellite retrieval or an objective analysis.
\newblock \textit{Journal of Applied Meteorology}, \textbf{32}, 1092--1107.

\bibitem[{Rabier et~al.(2002)Rabier, Fourri\'{e}, Chafai and
  Prunet}]{Rabier2002}
Rabier, F., Fourri\'{e}, N., Chafai, D. and Prunet, P. (2002) Channel selection
  methods for {Infrared Atmospheric Sounding Interferometer} radiances.
\newblock \textit{Quartely Journal of the Royal Meteorological Society},
  \textbf{128}, 1011--1027.

\bibitem[{Rodgers(2000)}]{Rodgers2000}
Rodgers, C. (2000) \textit{Inverse Methods for Atmospheric Sounding Theory and
  Practice}.
\newblock World Scientific Publising.

\bibitem[{Schraff and Coauthors(2016)}]{Schraff2016}
Schraff, C. and Coauthors (2016) {Kilometre-scale ensemble data assimilation
  for the COSMO model (KENDA)}.
\newblock \textit{Quartely Journal of the Royal Meteorological Society},
  \textbf{142}, 1453--1472.

\bibitem[{Tippett et~al.(2003)Tippett, Anderson, Bishop, Hamill and
  Whitaker}]{Tippett2003}
Tippett, M., Anderson, J., Bishop, C., Hamill, T. and Whitaker, J. (2003)
  Ensemble square root filters.
\newblock \textit{Monthly Weather Review}, \textbf{131}, 1485–1490.

\bibitem[{Tsyrulnikov(2010)}]{Tsyrulnikov2010}
Tsyrulnikov, M. (2010) {Is the Local Ensemble Transform Kalman Filter suitable
  for operational data assimilation?}
\newblock \textit{COSMO Newsletter}, \textbf{10}, 22--36.

\bibitem[{Wahba et~al.(1995)Wahba, Johnson, Gao and Gong}]{Wahba1995}
Wahba, G., Johnson, D., Gao, F. and Gong, J. (1995) {Adaptive Tuning of
  Numerical Weather Prediction Models: Randomized GCV in Three- and
  Four-Dimensional Data Assimilation}.
\newblock \textit{Monthly Weather Review}, \textbf{123}, 3358--3369.

\bibitem[{Wang et~al.(2004)Wang, Bishop and Julier}]{Wang2004}
Wang, X., Bishop, C. and Julier, S. (2004) Which is better, an ensemble of
  positive–negative pairs or a centered spherical simplex ensemble?
\newblock \textit{Monthly Weather Review}, \textbf{132}, 1590–1605.

\bibitem[{Wang et~al.(2007)Wang, Hamil, Whitaker and Bishop}]{Wang2007}
Wang, X., Hamil, T.~M., Whitaker, J.~S. and Bishop, C.~H. (2007) A comparison
  of hybrid ensemble transform {Kalman Filter–Optimum Interpolation} and
  ensemble square root filter analysis schemes.
\newblock \textit{Monthly Weather Review}, \textbf{135}, 1055--1076.

\bibitem[{Whitaker and Hamill(2002)}]{WhitakerHamill2002}
Whitaker, J.~S. and Hamill, T.~M. (2002) Ensemble data assimilation without
  perturbed observations.
\newblock \textit{Monthly Weather Review}, \textbf{130}, 1913–1924.

\bibitem[{Whitaker and Hamill(2012)}]{WhitakerHamill2012}
--- (2012) Evaluating methods to account for system errors in ensemble data
  assimilation.
\newblock \textit{Monthly Weather Review}, \textbf{140}, 3078–3089.

\bibitem[{Yamaguchi et~al.(2018)Yamaguchi, Hotta, Kanehama, Ochi, Ota,
  Sekiguchi, Shimpo and Yoshida}]{Yamaguchi18}
Yamaguchi, H., Hotta, D., Kanehama, T., Ochi, K., Ota, Y., Sekiguchi, R.,
  Shimpo, A. and Yoshida, T. (2018) Introduction to {JMA}'s new {Global
  Ensemble Prediction System}.
\newblock \textit{CAS/JSC WGNE Research Actitivies on Atmospheric and Oceanic
  Modelling}, \textbf{48}, 6.13--6.14.

\bibitem[{Zhang et~al.(2004)Zhang, Snyder and Sun}]{Zhang2004}
Zhang, F., Snyder, C. and Sun, J. (2004) {Impacts of initial estimate and
  observation availability on convective-scale data assimilation with an
  ensemble Kalman filter}.
\newblock \textit{Monthly Weather Review}, \textbf{132}, 1238–1253.

\end{thebibliography}


\clearpage
\section*{Appendix: DFS diagnostics applied to a quasi-operational global LETKF}
\setcounter{equation}{0}
\renewcommand{\theequation}{A\arabic{equation}}

\subsection*{Introduction}
The Degrees of Freedom for Signal (DFS), or analysis sensitivity to
observations, first introduced to NWP by \cite{Cardinali2004} and
\cite{Fisher2003}, is a convenient measure of how much of information
content a particular data assimilation (DA) system can extract from
different types of observations. Diagnostics of this quantity is useful
in identifying issues or limitations of a data assimilation system,
observation error specification, or of observations themselves. In this
Appendix, we briefly show the results from DFS diagnostics applied to a
quasi-operational version of global LETKF system developed at Japan
Meteorological Agency (JMA).

\subsection*{DFS calculation}
Following \cite{Liu2009}, DFS is calculated from the analysis
perturbations mapped onto the observation space, $\mathbf{Y}^a=\mathbf{HX}^a$, using 
\begin{align}
 \mbox{DFS}^\mathrm{ens}=\tr \mathbf{HK}
 =\tr \mathbf{H}\mathbf{A}^\mathrm{ens}\mathbf{H}^T\mathbf{R}^{-1}
 =\left(\mathbf{R}^{-1/2}\mathbf{Y}^a\right)^T\left(\mathbf{R}^{-1/2}\mathbf{Y}^a\right)/(K-1).
\end{align}
In our system, the prescribed observation error covariance matrix
$\mathrm{R}$ is chosen to be diagonal, so the DFS as calculated above
can be divided into contributions from each observation (which are just
the sample analysis variance corresponding to each observation normalized by
the corresponding observation error variance). Then, for each subset of
observations grouped by each instrument or each observed type, we can
define ``DFS per observation'' (which is also referred to as ``self
sensitivity''; see \cite{Cardinali2004} for detail). Monitoring of the
per-obs DFS thus defined for different observation types is a very
useful diagnostics.

\subsection*{Experimental setup}

The DA system analyzed here is a pre-operational version of JMA's global
LETKF that is operated to produce the initial perturbations used in the
global ensemble prediction system \citep{Yamaguchi18}. It is a 50-member
LETKF system, each member of which is a lower-resolution run (TL319 in
horizontal, $\sim 60$ km grid spacing, and 100 vertical levels reaching
from the surface up to 0.01 hPa) of JMA's Global Spectral Model
(GSM). At the end of each analysis update, the analysis mean is
recentered to the deterministic higher-resolution analysis produced by
4DVar. A detailed description of this LETKF system is given in section
3.3.3.1 of \cite{JMA-outline19}.

In this experiment, all the types of observations that are used
operationally by 4DVar are assimilated by the LETKF. The assimilated
observations are grouped into the following categories: SYNOP (surface
pressure measurements from ground-based stations), SHIP (surface
pressure measurements over the seas reported from vessels or moored
buoys), BUOY (as in SHIP but from drifting buoys), RADIOSONDE
(upper-level sounding observations of pressure, temperature, winds and
humidity reported from radiosondes), PILOT (upper-level wind
observations from rawin or pilot balloons), AIRCRAFT (aircraft
observations reported via AIREP or AMDAR programme), TYBOGUS (typhoon
bogus data), PROFILER (wind profiles measured from ground-based radars),
GNSS-DELAY (zenith total delay observations from ground-based GNSS
receivers), GNSS-RO (GNSS radio occultation observed by low earth orbit
satellites), AMVGEO (upper-level winds inferred as atmospheric motion
vectors from geostationary satellite imagery), AMVLEO (as in AMVGEO but
from lower earth orbit satellites), AMSU-A (microwave radiance soundings
from AMSU-A sensors), AIRS (hyper-spectral infrared sounding from AIRS
sensor), MHS (microwave humidity sounding from MHS sensors), IASI
(hyper-spectral infrared sounding from IASI sensor), SCATWIND (ocean
surface wind vectors inferred from ASCAT scatterometers), TMI (microwave
imagery from TMI sensor onboard TRMM satellite), AMSR2 (microwave
imagery from AMSR2 sensor onboard GCOM-W satellite), SSMIS (microwave
imagery from SSMIS sensors) and CSR (clear sky radiance imagery from
water-vapor-sensitive channels of geostationary satellites). The details
on these observations are documented in Section 2.2 of
\cite{JMA-outline13} or \cite{JMA-outline19}.

The results shown here are based on the statistics taken from the 5-day
period from 06 UTC, 10 July 2013 to 00 UTC, 15 July 2013. Just five days
may not be long enough for a reliable statistics, but we have confirmed
that the DFS statistics are not too different for different periods from
different seasons; to make the appendix concise, we only focus on this
particular 5-day period.

\subsection*{Overview of the results}

To gain insight as to which observation types are most informative in
terms of information content, we first examine the relative (fractional)
contributions from each observation type to the total DFS, defined as
the sum of DFS for each observation within each group divided by the
total DFS (the sum of DFS for all observations), which are plotted in
Figure \ref{fig:appendix-totalDFS}. The DFS is most contributed by
AMSU-A and GNSS-RO observations which together account for more than
half of the total DFS, followed by CSR, RADIOSONDE and AIRCRAFT.  We
highlight here that contributions from hyper-spectral soudings (AIRS and
IASI) are relatively small despite that they dominate in terms of the data
volume (the number of observations); this is in stark contrast to recent
DFS results from variational analysis at ECMWF \citep{Cardinali2013} and
M\'{e}t\'{e}o France \footnote{Real-time monitoring is available at
http://www.meteo.fr/special/minisites/monitoring/DFS/dfs.html.},
for instance, where a large contribution from hyper-spectral sounders
(notably IASI) is reported. 

The mean per-obs DFS, defined as $\tr \mathbf{HK}/p$ where $p$ is the
number of all the assimilated observations, is a measure of how much of
information the analysis extracts from observations on average. In some
literature \citep[e.g.][]{Cardinali2004,Lupu2011} this is called
Observation Influence (OI). The mean per-obs DFS for our LETKF system
was only 0.0157 (1.57\%), meaning, to the authors' surprise, that the
LETKF analysis relies about 98\% on the information from the background.
This is very small compared to results from variational systems. For
example, in a recent study from ECMWF's 4DVar, the mean per-obs DFS was
$\sim$ 11\% \citep{Cardinali2013}; similarly, \cite{Lupu2011} reported
that the mean per-obs DFS at Canadian operational 4DVar was about
10\%. In an idealized Observing System Simulation Experiment (OSSE)
using an intermediate global atmospheric model where rather sparse
in-situ observations are assimilated by LETKF \citep{Liu2009}, the
mean per-obs DFS was about 15\%. As we show in the next paragraph, the
smallness of the mean per-obs DFS can be attributed to the small per-obs
DFS for hyper-spectral sounders (AIRS and IASI) that constitute more than
70\% of the total observation count.

Figure \ref{fig:appendix-DFS-perobs} plots the per-obs DFS (or self
sensitivity) calculated for different types of observations, using
samples taken from (a) entire globe, (b) Tropics, (c) Northern
Hemisphere extratropics (NH), and (d) Southern Hemisphere extratropics
(SH).  In any of the regions, conventional (non-radiance) observations
tend to have higher per-obs DFS than satellite radiance observations
(except for CSR which are assigned relatively small error variance and
are relatively sparse compared to other radiance data due to the
cloud-freeness constraint and the strong horizontal thinning that picks
only one observation in a 200 km $\times$ 200 km box). Notably, isolated
observations like BUOY and SHIP exhibit large per-obs DFS. It can be
also observed that per-obs DFS for conventional observations (like
SYNOP, SHIP, BUOY and RADIOSONDE) are larger in SH and Tropics than in
NH. These features are consistent with \cite{Cardinali2004} who showed
that isolated observations tend to show larger DFS.

What is striking in Figure \ref{fig:appendix-DFS-perobs}, in comparison
to similar diagnostics from variational DA systems (e.g.,
\cite{Cardinali2013} and the real-time monitoring at M\'{e}t\'{e}o
France) is that per-obs DFS for hyper-spectral sounders (AIRS and IASI)
are very small. We discuss this point in the next subsection.

\subsection*{Discussion}

The formula for computing DFS applicable to any EnKF algorithm has been
devised in \cite{Liu2009} but appears not to have been applied to an
operational EnKF implementation. Here a DFS diagnostics is applied
perhaps for the first time to a quasi-operational global LETKF
implementation that assimilates all types of observations that are
operationally assimilated by 4DVar. 

The striking feature of the DFS diagnosed for LETKF, in comparison to
those reported for variational DA systems in the literature, is that the
mean per-obs DFS is very small. This feature is largely attributable to
the smallness of per-obs DFS for AIRS and IASI that constitute more than
70\% of the total data count. The important question then is to understand
why DFS is so small for these types of observations.

From the classical theory of DFS \citep{Cardinali2004,Fisher2003} (which
assumes that the Kalman gain $\mathbf{K}$ is accurately computed from
$\mathbf{B}$ and $\mathbf{R}$), in order for DFS to be small, the
observation error variance has to be large in comparison to its
counterpart from the background. In our LETKF, the observation error
covariance $\mathbf{R}$ is identical to what is used in the operational
4DVar, while the background error variance (as inferred from the
background ensemble spread) tends to be smaller than its counterpart
prescribed in 4DVar by a factor of $\sim 2^2$ or $3^2$. This discrepancy
in the magnitude of $\mathbf{B}$ is perhaps a factor contributing to the
discrepancies in DFS between LETKF and 4DVar in general, but it alone
cannot explain why per-obs DFS in LETKF are reasonably large for sparse
observations like BUOY and SHIP but are disproportionately small for
AIRS and IASI. It appears then that, to understand and explain the
discrepancy in DFS for AIRS and IASI between LETKF and 4DVar, we need to
examine how the Kalman gain $\mathbf{K}$ is computed in these two
algorithms. This question motivated the study presented in the main part
of this manuscript.

\clearpage
\section*{Figures}

\clearpage
\setcounter{figure}{0}
\renewcommand{\thefigure}{A\arabic{figure}}

\begin{figure}[htbp]
 \centering
 \includegraphics[bb=0 0 319 238, width=60mm]{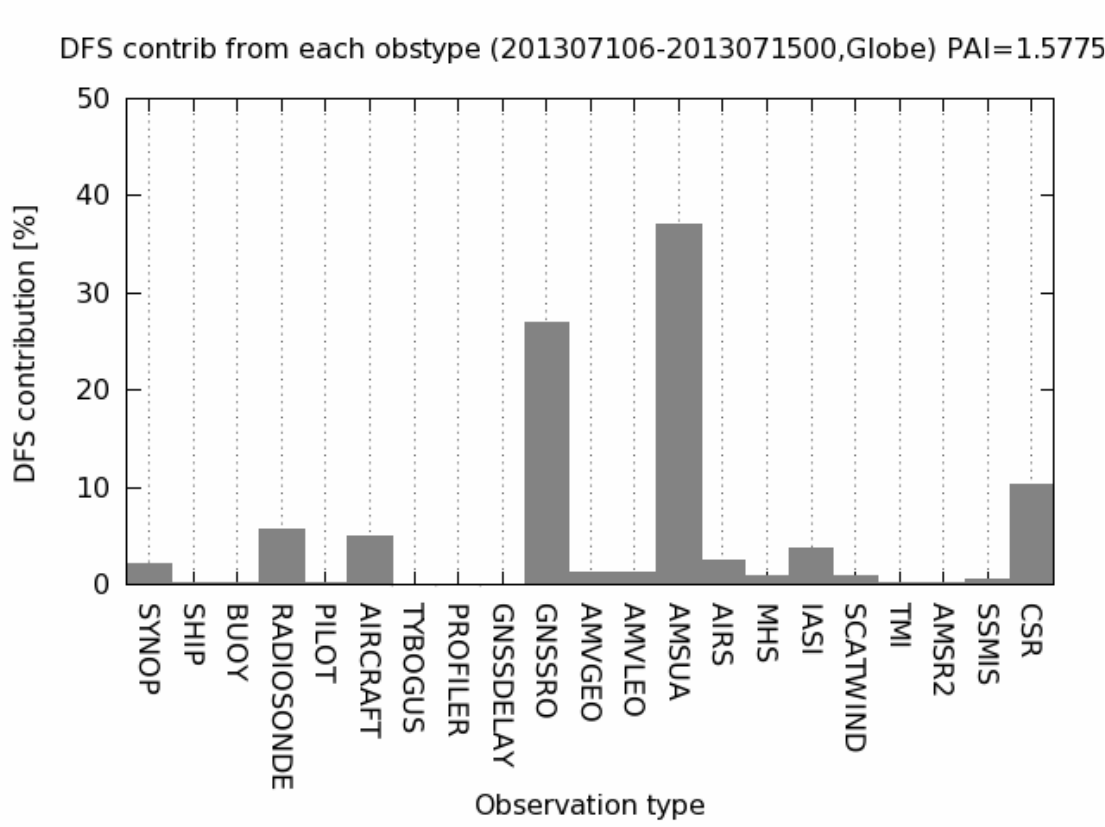} 
 \caption{
 Relative contributions to the total DFS from different types of observations.
 } \label{fig:appendix-totalDFS}
\end{figure}

\begin{figure}[htbp]
 \centering
 \includegraphics[bb=0 0 720 540, width=120mm]{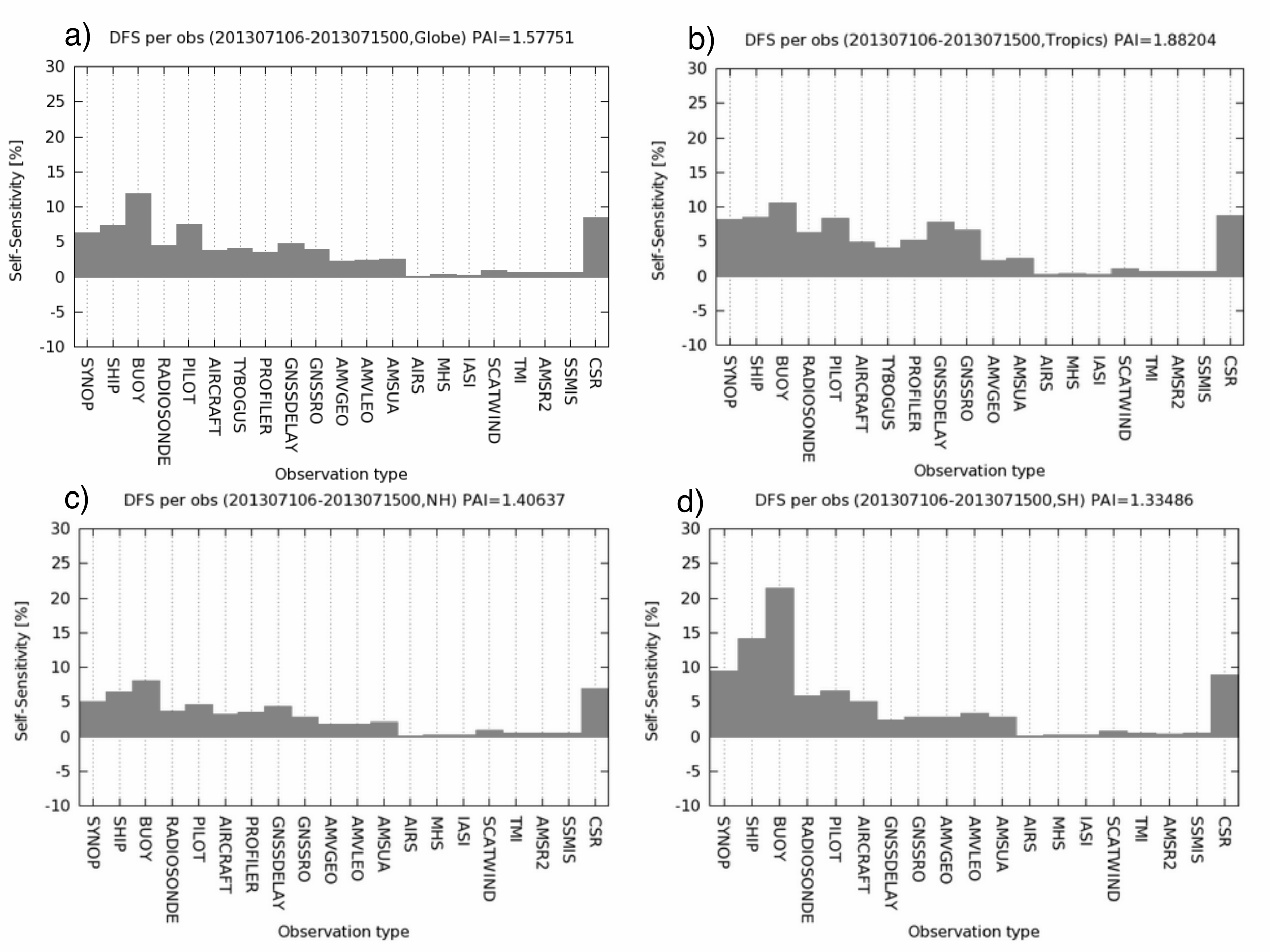} 
 \caption{
 DFS per observation (or self sensitivity) for different types of
 observations calculated using samples from different geographical
 areas. (a) the entire globe, (b) the Tropics
 (30{\textdegree}S--30{\textdegree}N), (c) Northern Hemisphere
 extratropics (30{\textdegree}N-90{\textdegree}N), and (d) Southern
 Hemisphere extratropics (90{\textdegree}S-30{\textdegree}S).
 } \label{fig:appendix-DFS-perobs}
\end{figure}


\clearpage

\end{document}